\documentclass[]{elsarticle} 
\usepackage{hyperref}
\usepackage{amssymb,amsfonts,amsmath,stmaryrd,amsthm,mathptmx}
\usepackage{bbm}
\usepackage{algorithm}
\usepackage{algorithmic}
\usepackage{subfigure}
\usepackage{xcolor}
\usepackage{float}
\usepackage{relsize}
\usepackage{fancyhdr}
\usepackage{makecell}
\journal{Computers and Geosciences}

\biboptions{authoryear}

\begin{document}

\begin{frontmatter}
    
    \title{Water Residence Time Estimation by 1D Deconvolution in the Form of a $l_{2}$-Regularized Inverse Problem With Smoothness, Positivity and Causality Constraints}
    
    
    \author{Alina G. Meresescu\fnref{fn1,fn2}}\ead{alina-georgiana.meresescu@u-psud.fr}
    \author{Matthieu Kowalski\fnref{fn2}}
    \author{Fr\'{e}d\'{e}ric Schmidt\fnref{fn1}}
    \author{Fran\c{c}ois Landais\fnref{fn1}}
    
    \fntext[fn1]{GEOPS, Univ. Paris-Sud, CNRS, Universite Paris-Saclay, Rue du Belvedere, Bat. 504-509, 91405 Orsay, France}
    \fntext[fn2]{L2S, Univ. Paris-Sud, Supelec, CNRS, Universite Paris-Saclay, 3 rue Joliot Curie, 91192 Gif-sur-Yvette, France}
    
    
    \cortext[mycorrespondingauthor]{Corresponding author}

    
    \begin{abstract}
        The Water Residence Time distribution is the equivalent of the impulse response of a linear system allowing the propagation of water through a medium, e.g. the propagation of rain water from the top of the mountain towards the aquifers. We consider the output aquifer levels as the convolution between the input rain levels and the Water Residence Time, starting with an initial aquifer base level. The estimation of Water Residence Time is important for a better understanding of hydro-bio-geochemical processes and mixing properties of wetlands used as filters in ecological applications, as well as protecting fresh water sources for wells from pollutants. Common methods of estimating the Water Residence Time focus on cross-correlation, parameter fitting and non-parametric deconvolution methods. Here we propose a 1D full-deconvolution, regularized, non-parametric inverse problem algorithm that enforces smoothness and uses constraints of causality and positivity to estimate the Water Residence Time curve. Compared to Bayesian non-parametric deconvolution approaches, it has a fast runtime per test case; compared to the popular and fast cross-correlation method, it produces a more precise Water Residence Time curve even in the case of noisy measurements. The algorithm needs only one regularization parameter to balance between smoothness of the Water Residence Time and accuracy of the reconstruction. We propose an approach on how to automatically find a suitable value of the regularization parameter from the input data only. Tests on real data illustrate the potential of this method to analyze hydrological datasets.
    \end{abstract}
    
\end{frontmatter}


\section{Introduction}

The hydrological \textit{Water Residence Time distribution} (named in this article simply as residence time) is a measure allowing the analysis of the transit of water through a given medium. Its estimation is necessary when using wetlands as a natural treatment plant for pollutants that are already in the water~\cite{Z_Hydro_WernerKadlec2000}, to better manage and protect drinking water sources from pollution~\cite{Z_Hydro_Cirpka2007}, to study the water transport of dissolved nutrients~\cite{Z_Hydro_Gooseff2011}. For a more comprehensive application range, including deciphering hydro-bio-geochemical processes or river monitoring, the review done in~\cite{Z_Hydro_McGuireMcDonnell2006} is a useful starting point. We call here the residence time the linear response of the aquifer system. In this context it refers to wave propagation of the water dynamics, not to the actual molecular travel time~\cite{Z_Hydro_Botter2011}.

To obtain the residence time, one can distinguish two families of methods: active and passive. The active methods are carried out by releasing tracers at the entrance of the system at a given time, like artificial dyes, and then by tracing the curve while measuring the tracer levels at the exit of the system~\cite{Z_Hydro_DZIKOWSKI1992697, Z_Hydro_WernerKadlec2000, Z_Hydro_PaynGooseff2008, Z_Hydro_Robinson2010}. Although robust, this methodology  involves high effort and high operational costs. It could also perturb the water channel and this may lead to biased results. The passive methodology consists of recording data at the inlet and outlet of the water channel by specific water isotopes~\cite{Z_Hydro_McGuireMcDonnell2006}, water electrical conductivity~\cite{Z_Hydro_Cirpka2007} or by simply recording the rainfall levels at high altitude grounds and the aquifer levels at the base ~\cite{Z_Hydro_Delbart_2014}. In the passive case, the residence time is not measured directly but must be retrieved by deconvolution. Some authors also use deconvolution in the active methodology when the release of tracer cannot be considered as instantaneous~\cite{Z_Hydro_McGuireMcDonnell2006, Z_Hydro_Cirpka2007, Z_Hydro_PaynGooseff2008}. The residence time can then be approximated as the impulse response of the system and this in turn can be estimated by deconvolution~\cite{Z_Hydro_Neuman1982, Skaggs1998, Fienen2006}. The method can also be used for enhancing geophysical models, although not targeted explicitly for Water Residence Time estimation~\cite{Z_Hydro_Zuo2012170}. Deconvolution methods can be parametric~\cite{Z_Hydro_NeumanDeMarsily1976, Long1999, Z_Hydro_Etcheverry2000, Z_Hydro_WernerKadlec2000, Luo2006, Z_Hydro_McGuireMcDonnell2006} or non-parametric~\cite{Z_Hydro_Neuman1982, Dietrich1993, Skaggs1998, Michalak2003, Z_Hydro_Cirpka2007, Fienen2008, Z_Hydro_Gooseff2011, Z_Hydro_Delbart_2014}.

Parametric methodology has the advantage of always providing a result with expected properties such as correct shape and positiveness but with the caveat of being insensitive to unexpected results for real data (for instance a second peak in the residence time). The non-parametric deconvolution has the advantage of being "blind", meaning that no strong \textit{a priori} are being set on the estimated curve, but in the absence of adapted mathematical constraints, the results may not reflect the physics of the residence time curve (these are sometimes negative or non-causal). 

Our method is non-parametric and takes into account limitations of previous methods from the same category: variable-sized rainfall time series as input compared to~\cite{Z_Hydro_Neuman1982}, a more compact direct model formulation than in~\cite{Z_Hydro_Neuman1982, Z_Hydro_Cirpka2007}, less computational effort and less time consuming than for a Bayesian Monte-Carlo inverse problem methodology~\cite{Fienen2006, Fienen2008},  strictly using a passive method with respect to mixed methods like the ones in~\cite{Z_Hydro_Gooseff2011}. In contrast to the cross-correlation~\cite{ Vogt2010, Z_Hydro_Delbart_2014} we avoid the unrealistic hypothesis that the rain signal can be considered as white noise. In fact, rainfall datasets have long range memory properties and therefore we simulate the input rainfall for synthetic tests as a multifractal signal~\cite{Z_Hydro_Tessier1996b}. One important difference from other non-parametric deconvolution methods is that we enforce causality explicitly through projection. We also discuss the importance of this aspect to avoid a sub-optimal solution when using a Fourier Domain based convolution \cite{McCormick1969}. In~\cite{Z_Hydro_Neuman1982, Dietrich1993, Z_Hydro_Delbart_2014} the causality constraint was not mentioned. In~\cite{ Skaggs1998, Z_Hydro_Cirpka2007, Z_Hydro_PaynGooseff2008, Z_Hydro_Gooseff2011}, causality is taken into account through a carefully constructed Toeplitz matrix for the convolution operation.

We propose a new algorithm to estimate the residence time with the following properties:
\begin{itemize}
    \item passive: only input rainfall and output aquifer levels are required;
    \item flexible: in the sense that it handles even unexpected solutions (double peaks or unexpected shapes of the residence time). It can handle Dirac-like rain events as inputs but also clustered rain events over a longer time period (for instance a whole season);
    \item constrained: by physical and mathematical aspects of the residence time (positivity, smoothness and causality);
    \item automatic: providing a simple and accurate way of choosing the best hyper-parameter that governs the smoothness of the residence time curve, without human operation;
    \item efficient/accurate: a fast algorithm that provides a good signal-to-noise ratio (SNR), avoiding noise amplification.
\end{itemize}
This last property is important in order to deal with  non-linearity and non-stationarity of the water channel, a known difficulty in residence time estimation~\cite{Z_Hydro_NeumanDeMarsily1976, Z_Hydro_Massei2006, Z_Hydro_McGuireMcDonnell2006, Z_Hydro_PaynGooseff2008} 

The rest of this article is organized as follows: Section~\ref{sec:model} presents the direct problem and the inverse problem formulation, Section~\ref{sec:AM} depicts the algorithm used to solve this inverse problem formulation. Some important implementation details are discussed in Section~\ref{sec:implementation}. We also discuss differences between our solution and previous non-parametric 1D deconvolution methods used as benchmarks in Section~\ref{sec:discussion}. In Section~\ref{sec:synthetic} we present results obtained from synthetic data and we discuss the choice of the hyper-parameter that controls the smoothness of the residence time. Finally, we present results obtained from real data in Section~\ref{sec:real}, while Section~\ref{sec:conclusion} concludes the paper.

\section{Model}
\label{sec:model}

\subsection{Direct Problem}

The direct model for water propagation through a channel can be written as a linear system~\cite{Z_Hydro_Neuman1982}:
\begin{equation}\label{1}
\begin{aligned}
\textbf{y} = \mathbbm{1}c + \textbf{x} \ast \textbf{k} + \textbf{n}\ ,
\end{aligned}
\end{equation}
with:
\begin{itemize}
    \item $\textbf{y} \in \mathbbm{R_+}^{T}, \textbf{y} = (y_{0}, ... , y_{T})$ output of the linear system: aquifer basin level (known), real, positive signal, of length $T$,
    \item $\mathbbm{1}$ vector of all ones of length $T$,
    \item $c\geq 0$ initial aquifer basin level (to estimate), real, positive, constant
    \item $\textbf{x} \in \mathbbm{R_+}^{T}, \textbf{x} = (x_{0}, ... , x_{T})$  input of the linear system: rainfall level (known), real, positive signal, of length $T$
    \item $\ast$ convolution
    \item $\textbf{k} \in \mathbbm{R_+}^{K}, \textbf{k} = (k_{-\frac{K}{2}}, ... k_{0}, k_{1}, ... k_{\frac{K}{2}})$  impulse response to be estimated, real, positive signal, of length $K$
    \item $\textbf{n}\in\mathbbm{R}^T$  white gaussian noise, real, signal of length $T$.
\end{itemize}

The impulse response of the system -- $\textbf{k}$ -- as well as the mean level of the aquifer -- $c$ -- must be estimated. It is required that $\textbf{k}$ be positive, causal, and smooth. If positivity is obvious for the residence time, causality refers to the delayed, unidirectional flow of water from the point of entry to the aquifer, thus the idea that  $\textbf{k}$ must progress only in the positive time domain (negative time domain elements of $\textbf{k}$ are zero). Smoothness regularization is used in order to avoid noise amplification in the deconvolution.

\subsection{Inverse Problem}

To estimate $\textbf{k}$, we propose to solve the following constrained optimization problem:
\begin{align}
\label{2}
\hat{\textbf{k}},  \hat{c} & = \underset{\textbf{k}\in\mathbbm{R}_+^K,c}{\text{argmin}}  \frac{1}{2} \|\textbf{y} - \textbf{x} \ast \textbf{k} - c\mathbbm{1} \|_2^2 + \lambda \|\nabla \textbf{k}\|_{2}^{2}  \\
s.t. & \quad  \text{causality is enforced:}\quad \forall i\in\{-K/2,\ldots,-1\}\ k_i = 0
\nonumber
\end{align}

This function classically introduces a "fidelity term" (attachment to the data) corresponding to the white Gaussian  noise, as well as a $\ell_2$ "regularization term" on the gradient of $\textbf{k}$ in order to favor "smooth" solutions. The smoothness degree of the estimate is controlled by the hyper-parameter $\lambda$. A bigger $\lambda$ will stress more the smoothness of the solution, while a smaller $\lambda$ will better fit the solution to the data. A main goal of this work is also to find the optimal $\lambda$ range that consistently gives accurate estimates while taking into account both good data representation and smoothness \textit{a priori}. In the following, we rewrite the functional~\eqref{2} using matrix operators:
\begin{align}
\label{3}
J(\textbf{k},c) & =   \frac{1}{2} \|\textbf{y} - \textbf{X} \textbf{k} - c\mathbbm{1} \|_2^2 + \lambda \|\textbf{D} \textbf{k}\|_{2}^{2}  \\
s.t. & \quad  \forall i\in\{-K/2,\ldots,-1\}\ k_i = 0\text{ and } \forall i\ k_i \geq 0
\nonumber
\end{align}

where $\textbf{X}$ is the Toeplitz  matrix corresponding to the convolution by the signal $\textbf{x}$, while  $\textbf{D}$ is the finite-difference matrix corresponding to the gradient used for applying smoothness on the estimated signal. The minimization of $J(\textbf{k},c)$ can be interpreted as a Maximum A Posteriori (MAP) estimation in a Bayesian context with a Gaussian prior on the noise and an exponential family on the smoothness.

Since the problem is convex, we estimate $\textbf{k}$ and $c$ by an Alternating Minimization algorithm (shortened throughout as AM), that ensures a global minimization for the two items to be estimated. A historical overview is available from~\cite{Z_Hydro_O_Sullivan_1998}. With a fixed $c$, the problem is a simple quadratic optimization with constraints that is solved using the Projected Newton Method~\cite{Z_Hydro_bertsekas1982}, chosen for computational speed. With a fixed $\textbf{k}$, the estimate of $c$ is given by an analytic formula.

The AM algorithm will evaluate $\textbf{k}$ to convergence while applying an orthogonal projection $P$ on the positivity and causality constraints in each iteration. The analytic solution for $\textbf{k}$ is computed and used as an initial step for the iterative AM algorithm.

\section{Alternating Minimization for 1D Deconvolution}
\label{sec:AM}

After replacing the convolution operator with the equivalent Toeplitz matrix $\textbf{X}$, we introduce the functional $J(\textbf{k}, c)$ to minimize:
\begin{equation}\label{4}
\begin{aligned}
& J(\textbf{k},c) =  P \left( \dfrac{1}{2} || \textbf{y} - \textbf{X} \cdot \textbf{k} - c \mathbbm{1}||_{2}^{2} + \lambda ||\textbf{Dk}||_{2}^{2} \right)\ ,
\end{aligned}
\end{equation}
where $ P(\textbf{k}) \text{  is the orthogonal projection over the constraints, } \forall t \text{   }  \textbf{k}_{t} = 0 \text{ if }  \textbf{k}_{t} < 0 \text{ or if } t<0$.

Considering that both $\textbf{k}$ and $c$ must be estimated, we propose an AM algorithm where in a first step $\textbf{k}_{est}$ is estimated, then in the second step $c_{est}$ is updated.

\subsection{Estimation of \texorpdfstring{$\textbf{k}_{est}$}{k \textsubscript{est}} with the Projected Newton Method}

The update of $\textbf{k}_{est}$  by the Projected Newton Method with $c$ fixed is given by:
\begin{align}
\label{5}
\nonumber
\textbf{k}_{t+1} & = P\left( \textbf{k}_{t} + \alpha_{t} \cdot (- \nabla^{2} J(\textbf{k},c) ^{-1} \cdot  \nabla J(\textbf{k},c) ) \right) \\
& = P\left( (1- \alpha_{t}) \textbf{k}_{t} + \alpha_{t} \cdot (\textbf{X}^{T}\textbf{X} + \lambda \textbf{D}^{T}\textbf{D})^{-1} \cdot \textbf{X}^{T}\left(\textbf{y}-c\mathbbm{1}\right)  \right)\ ,
\end{align}
where $\alpha_t>0$ is the descent step size.
For $\textbf{k} = \left\{k_{-K/2},\ldots,k_0,k_{K/2}\right\}$, we have $P(\textbf{k}) = \left\{0,\ldots,0,(k_0)^+,\ldots,(k_{K/2})^+\right\}$, where $(x)^+=\max(0,x)$.

By replacing the Hessian and the Jacobian of \eqref{3} in \eqref{5}, we see that only the step size $\alpha_t$ can evolve at each iteration, while $\textbf{k}_{t}$ is changed by a constant called Newton's step.
\begin{equation}
\begin{aligned}\label{6}
&\textbf{k}_{t+1} = (1- \alpha_{t}) \textbf{k}_{t} + \alpha_{t} \cdot (\textbf{X}^{T}\textbf{X} + \lambda \textbf{D}^{T}\textbf{D})^{-1} \cdot \textbf{X}^{T}\tilde{\textbf{y}}\\
&\textbf{k}_{t+1} = (1- \alpha_{t}) \textbf{k}_{t} + \alpha_{t} \cdot \vartriangle_{n}^{t}\ ,
\end{aligned}
\end{equation}
where $\alpha_{t}$ is the  variable step size,
$ \vartriangle_{n}^{t} = (\textbf{X}^{T}\textbf{X} + \lambda \textbf{D}^{T}\textbf{D})^{-1} \cdot \textbf{X}^{T}\tilde{\textbf{y}}$ is Newton's step.

\subsection{Estimation of \texorpdfstring{$c$}{c}}

Taking the derivative of~\eqref{3} with respect to $c\mathbbm{1}$ leads to:
\begin{equation}\label{7}
\begin{aligned}
&\nabla J(\textbf{k},c) = - \textbf{y} + \mathbbm{1}^{T}\textbf{X} \textbf{k} + c\mathbbm{1} \overset{!}{=} 0\\
\end{aligned}
\end{equation}

With $\textbf{k}$ fixed, the estimation of $c$ is given by:
\begin{equation}\label{8}
\begin{aligned}
&c\mathbbm{1} =\overline{\textbf{y} - \mathbbm{1}^{T}\textbf{X} \textbf{k}}\ ,
\end{aligned}
\end{equation}
where $\bar{m}$ is the empirical mean of vector $m$.

The AM algorithm for estimating $\textbf{k}$ and $c$ is summarized in Alg.~\ref{alg_ma}.

\begin{algorithm}[H]
    \caption{Alternating Minimization}
    \label{alg_ma}
    \begin{algorithmic}[1]
        \renewcommand{\algorithmicrequire}{\textbf{Input:}}
        \renewcommand{\algorithmicensure}{\textbf{Output:}}
        \REQUIRE $\textbf{x},\textbf{y}, \lambda, D, \alpha_{min}, k\_err_{min}, y\_err_{min}, s_{max}, t_{max}$
        \ENSURE  $\textbf{k}_{est}$,$c_{est}$,$\textbf{y}_{rec}$\\
        \STATE  $c_{est} = \overline{\textbf{y}}, \hat{\textbf{y}} = \textbf{y} - c_{est}$
        \STATE $\vartriangle_{n}^{t} = (\textbf{X}^{T}\textbf{X} + \lambda \textbf{D}^{T}\textbf{D})^{-1} \cdot \textbf{X}^{T}\hat{\textbf{y}}$,   $\textbf{k}_{est} = \vartriangle_{n}^{t}$
        \STATE $k\_err_{rel} = 1, y\_err_{rel} = 1, s = 0, t = 0$,  $J_{ref} = \dfrac{1}{2} ||\hat{\textbf{y}}||^{2}$, $\textbf{y}_{rec} = \mathbbm{1}$
        \WHILE {$s$ != $s_{max}$ and $y\_err_{rel}> y\_err_{min}$}
        \STATE $\alpha = 1$, $s = s + 1$
        \STATE $\textbf{k}_{est\_old} = \textbf{k}_{est}$, $\textbf{y}_{rec\_old} = \textbf{y}_{rec}, \hat{\textbf{y}} = \textbf{y} - c_{est}$
        \WHILE {$t$ != $t_{max}$ and $k\_err_{rel}>k\_err_{min}$ and $\alpha> \alpha_{min}$}
        \STATE $t = t + 1$
        \STATE $\textbf{k}_{est} = P \left((1-\alpha)\textbf{k}_{est\_old} + \alpha  \vartriangle_{n}^{t}\right)$
        \STATE $J(k)^{t+1} = \dfrac{1}{2} ||\hat{\textbf{y}} - \textbf{x} \ast \textbf{k}_{est}||_{2}^{2} + \lambda ||D \textbf{k}_{est}||_{2}^{2} $
        \IF {($J(k)^{t+1} > J_{ref}$)}
        \STATE $\textbf{k}_{est\_old} = \textbf{k}_{est}$, $\alpha = 0.9 \cdot \alpha$
        \ELSE
        \STATE $J_{ref} = J(k)^{t+1}, t = 0$
        \STATE break;
        \ENDIF
        \STATE $k\_err_{rel} = \dfrac{||\textbf{k}_{est} - \textbf{k}_{est\_old}||_{2}^{2}}{||\textbf{k}_{est}||_{2}^{2}}$
        \ENDWHILE
        \STATE $\tilde{\textbf{y}}_{rec} = \textbf{x} \ast \textbf{k}_{est}$
        \STATE $c_{est} = \overline{\textbf{y} -  \tilde{\textbf{y}}_{rec}}$
        \STATE $\textbf{y}_{rec} =  \tilde{\textbf{y}}_{rec} + c_{est}$, $y\_err_{rel} = \dfrac{||\textbf{y}_{rec} - \textbf{y}_{rec\_old}||_{2}^{2}}{||\textbf{y}_{rec}||_{2}^{2}}$
        \ENDWHILE
        \RETURN $k_{est}$, $y_{rec}$, $c_{est}$
    \end{algorithmic}
\end{algorithm}

\section{Implementation Details}
\label{sec:implementation}

We provide a distribution package in Matlab for our algorithm and the download link can be found at the end of this article.
Although in the previous sections the model and the solution are written in matrix form, the Matlab implementation of the convolution for our AM algorithm is done through dot product multiplication in the Fourier Domain with appropriate zero padding,  meaning that no Toeplitz matrix is explicitly defined here for the convolution. It is also possible to carefully implement a causal convolution by designing a proper Toeplitz matrix. However, the convolution in the Fourier Domain appears to be more efficient in general.

This implementation also allows for the estimation of a $\textbf{k}$ residence time longer than the inputs $\textbf{x}$ and $\textbf{y}$, although this would be under-determined. Once that non-circularity is enforced through this particular implementation of the convolution, another aspect that is dealt with is the causality constraint.

\begin{figure}[H]
    \centering
    \subfigure[]{\label{subfig_1}\includegraphics[clip,trim=2.3cm 1.7cm 2.7cm 1.8cm,scale=0.32]{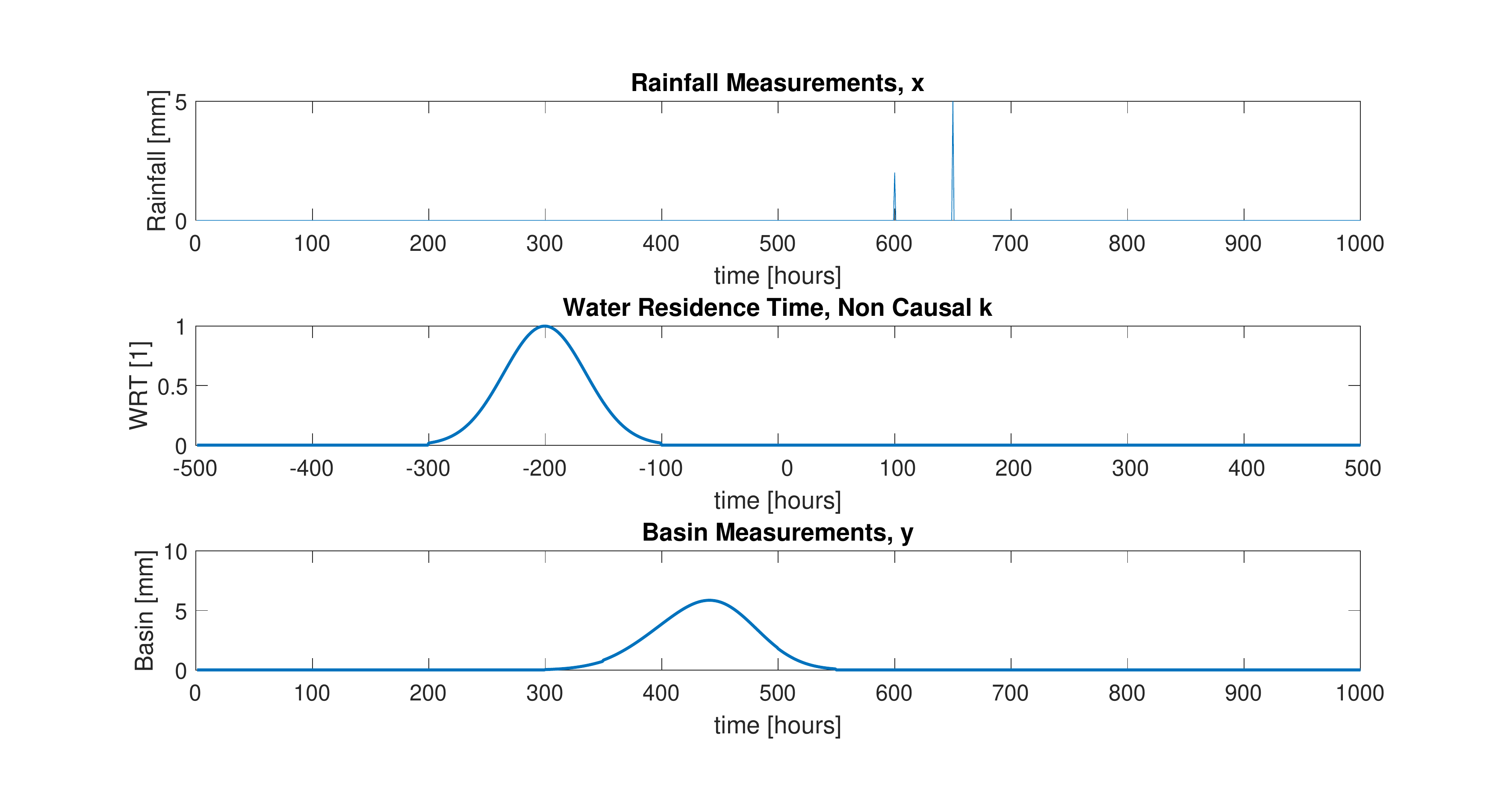}}\\
    \subfigure[]{\label{subfig_2}\includegraphics[clip,trim=2.3cm 1.7cm 2.7cm 1.8cm,scale=0.32]{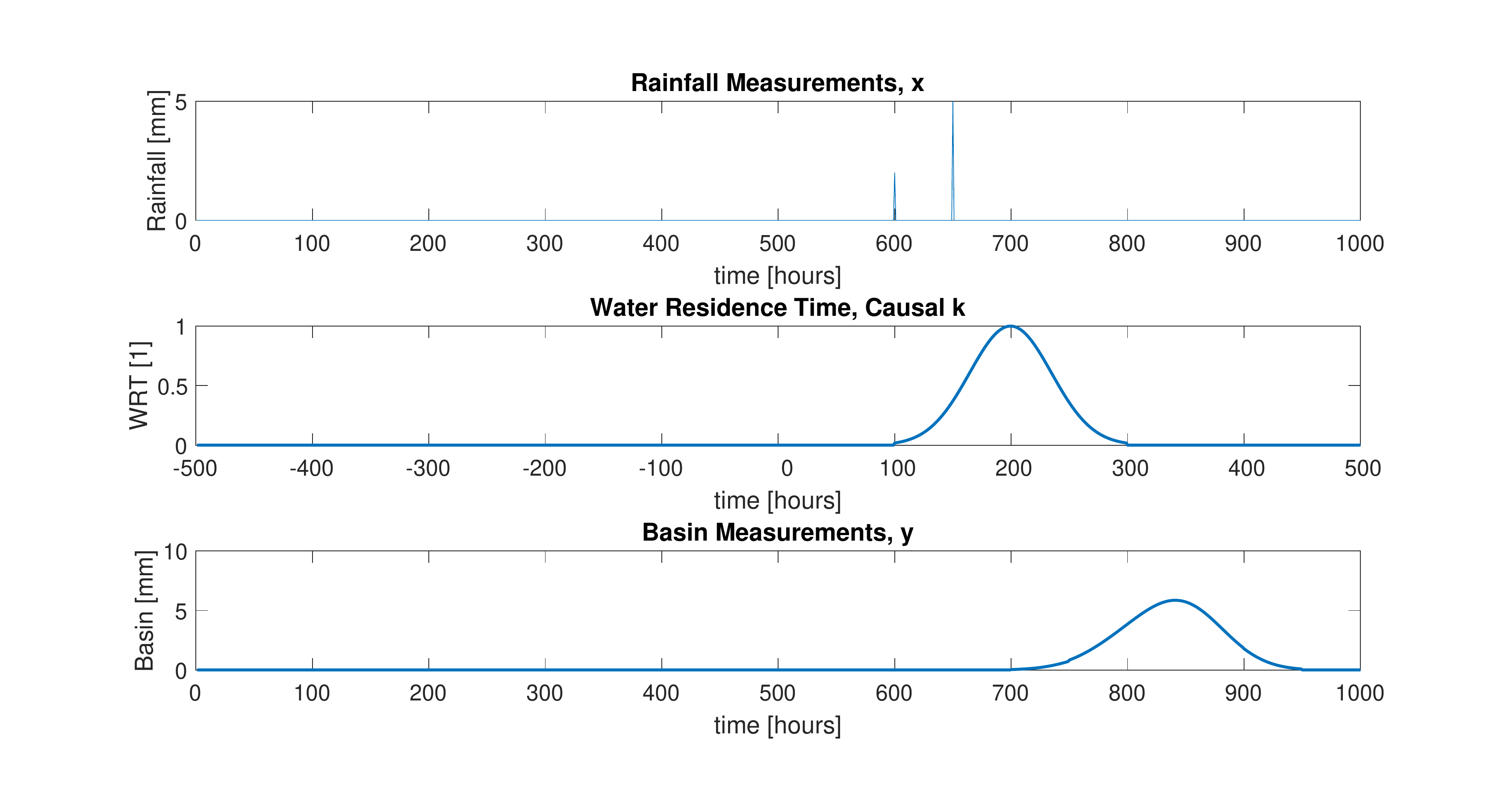}}
    \caption{Enforcing causality while doing the convolution in the Fourier Domain needs to include the negative time domain interval of the residence time.}
    \label{fig_causal_nocausal}
\end{figure}

In Figure~\ref{fig_causal_nocausal}, we present the convolution of two rainfall Diracs with a residence time curve. We convolve the rainfall time series once with a residence time curve found in the negative time domain (causality is not respected) and once when this curve is in the positive time domain (causality is respected). The resulting breakthrough curve appears before the rain events in the first case which is wrong. In the second case the breakthrough curve appears after these rainfall events as expected for real applications. In the non-causal case lobes can appear in the negative time domain also, incorporating energy that should be present in the residence time curve thus reducing its amplitude and distorting its shape.

\begin{figure}[H]
    \centering
    \subfigure[]{\label{subfig_3}\includegraphics[clip,trim=2.3cm 1.3cm 2.7cm 1.8cm,scale=0.3]{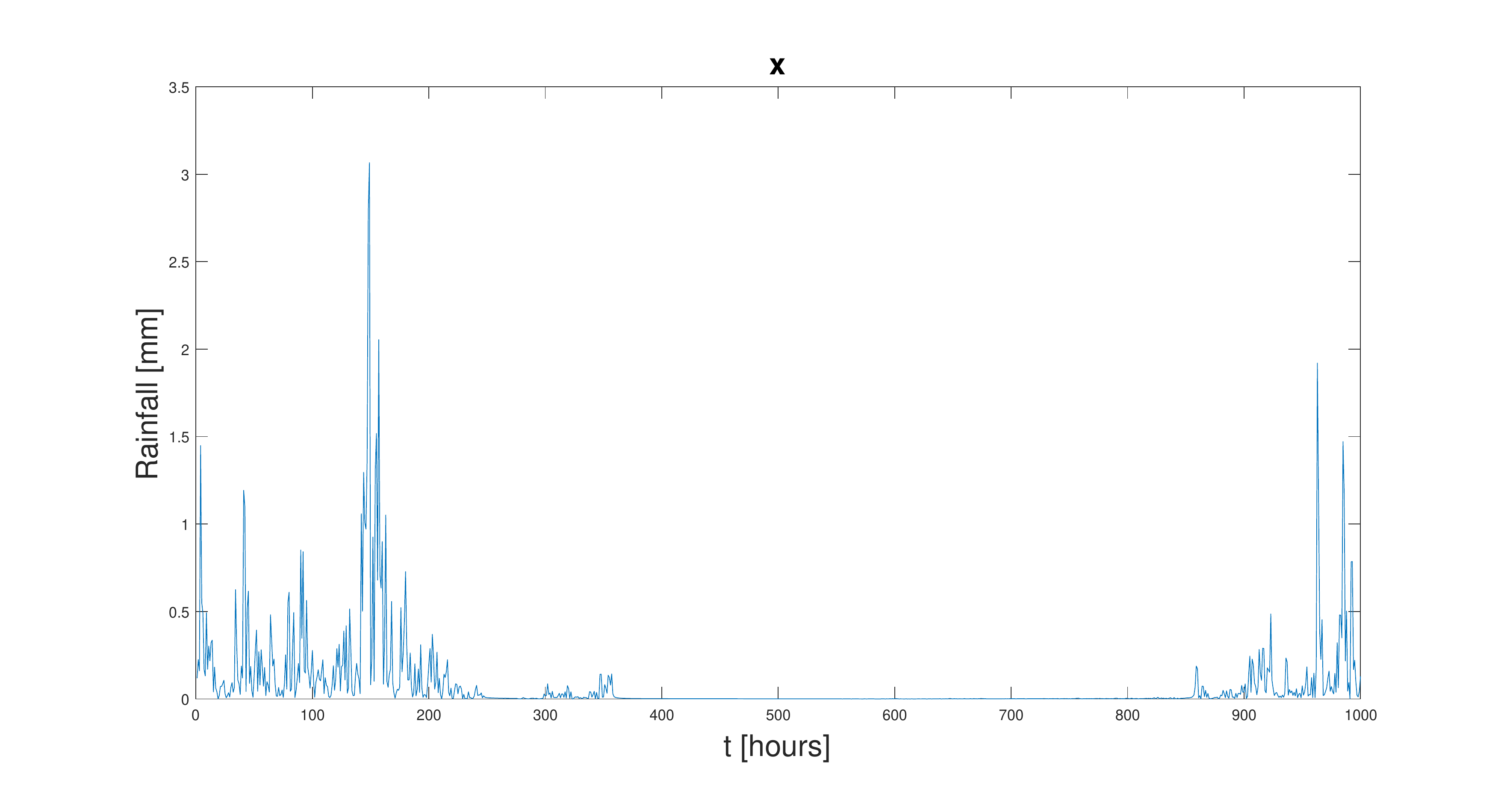}}
    \subfigure[]{\label{subfig_4}\includegraphics[clip,trim=2.3cm 1.3cm 2.7cm 1.8cm,scale=0.3]{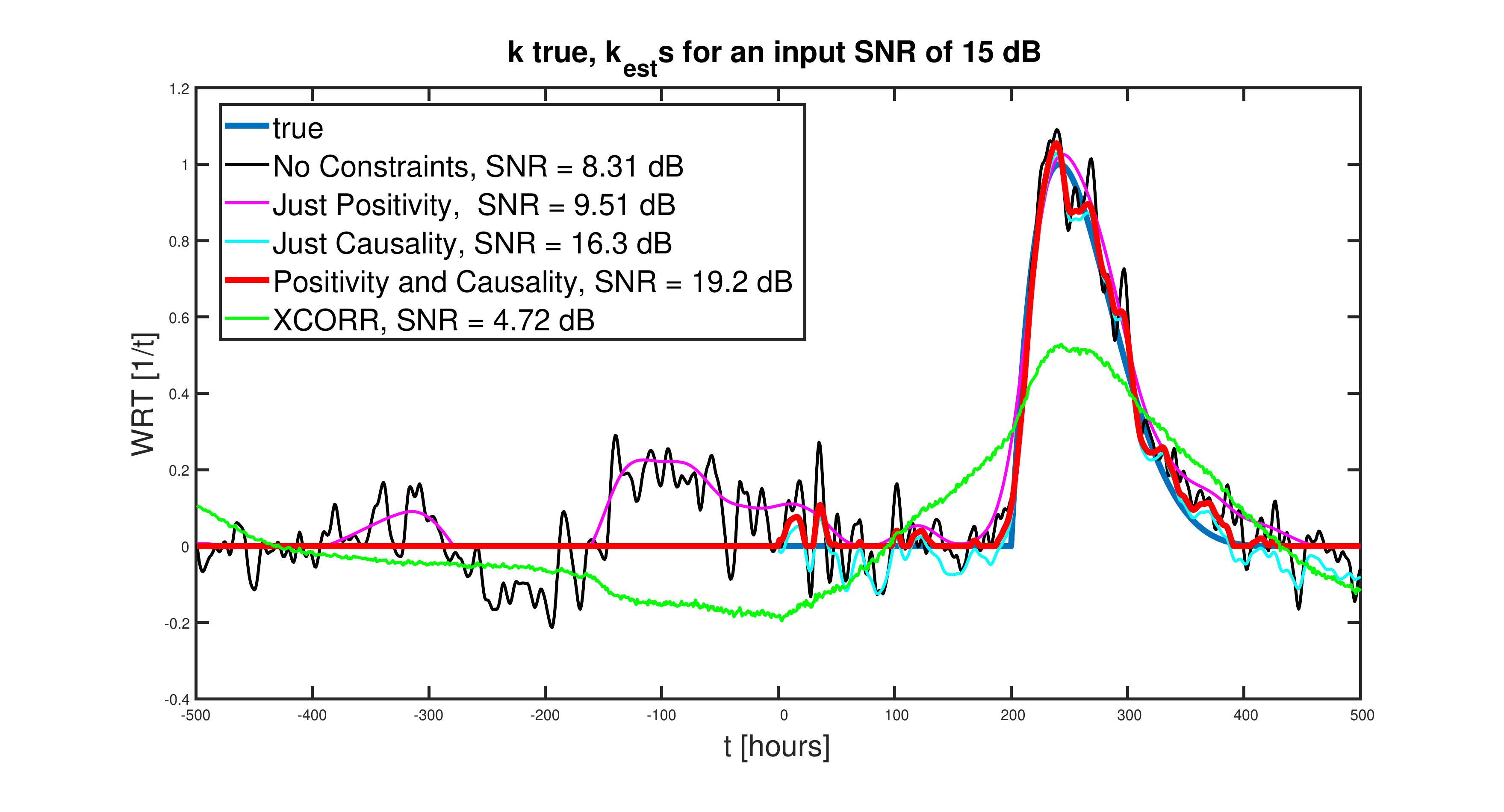}}
    \subfigure[]{\label{subfig_5}\includegraphics[clip,trim=2.3cm 1.3cm 2.7cm 1.8cm,scale=0.3]{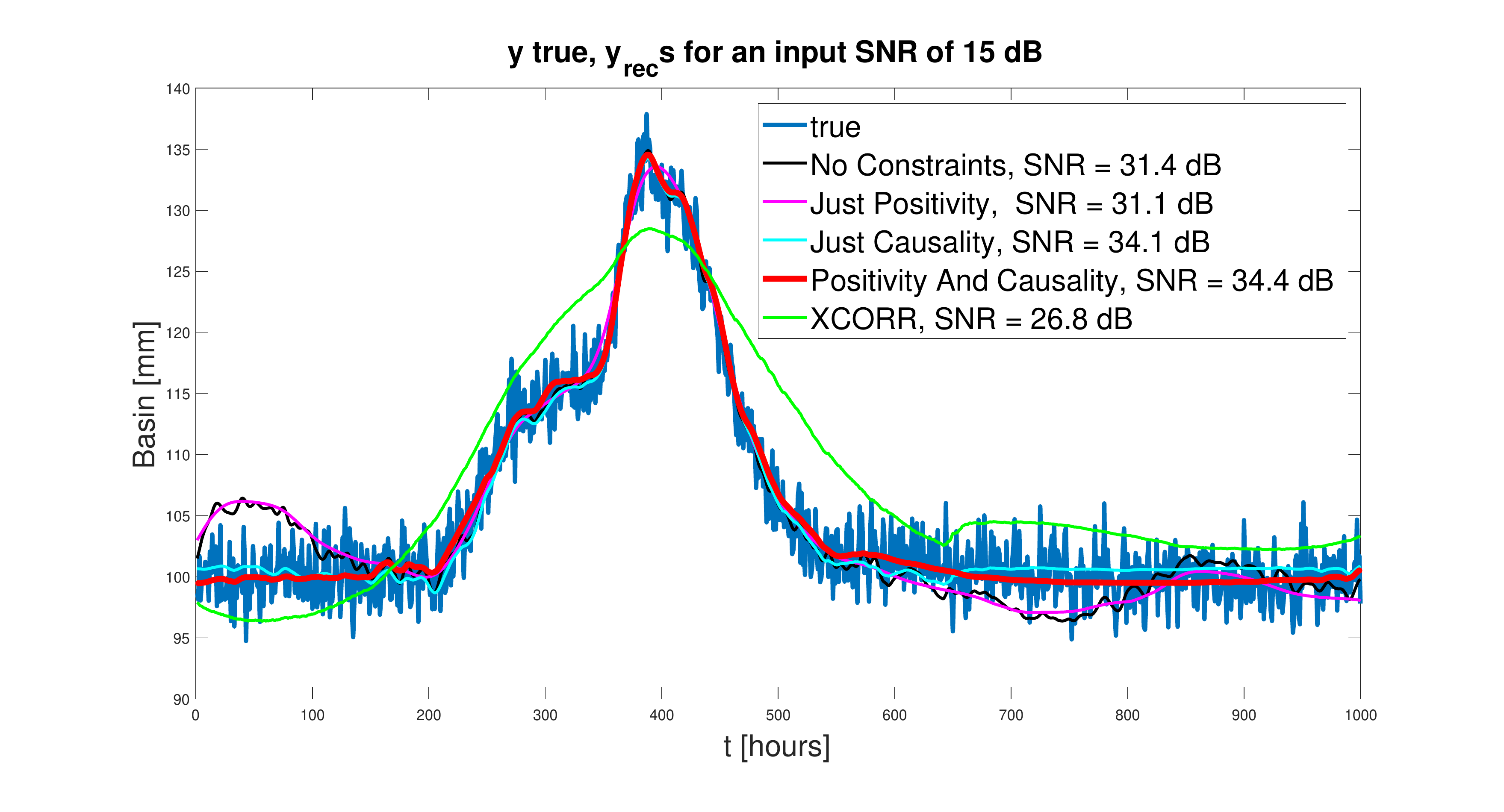}}
    \caption{ Different results for the $\textbf{k}_{est}$ for different constraints applied during the AM algorithm. All give a similar $\textbf{y}_{rec}$ but the best  $\textbf{y}_{rec}$ and $\textbf{k}_{est}$ are those where both positivity and causality constraints are applied. }
    \label{fig_full_k}
\end{figure}

In Figure~\ref{fig_full_k} we estimate with the AM algorithm all the possible residence time curves: with no positivity and no causality constraints applied, only the positivity constraint applied, only the causality constraint applied, and both positivity and causality constraints applied. In all cases, the convolution between the rainfall and these residence time curves give a reconstructed breakthrough curve that is similar in general shape with the real one. The best residence time estimation and breakthrough curve reconstruction are nonetheless the ones where both positivity and causality constraints are applied in the algorithm.

Furthermore, not applying the causality constraint all along the AM algorithm, and  setting the negative time domain of $\textbf{k}_{est}$ to zero only at the end, would lead to a suboptimal solution caused by the way in which the AM algorithm navigates through the optimality map attached to the given functional: any change in the estimated vector $\textbf{k}_{est}$ at the end of the algorithm moves the value of the functional away from the optimal point that was estimated in the last iteration~\cite{McCormick1969, Z_Hydro_bertsekas1982}.

\section{Discussion on Related Work}
\label{sec:discussion}
Non-parametric deconvolution techniques with/without positivity constraints exist from the 1980s. How is our method different from those and why benchmarking it against the cross-correlation?

\subsection{Comparison to Previous Works}

As a first example, let's take~\cite{Z_Hydro_Neuman1982} which does a regularized non-parametric deconvolution and uses a bi-criterion curve; it navigates the optimality map to find the optimal estimation of the residence time by using a lag-one auto-correlation coefficient between the two error criteria. We consider this to be similar to our approach but our functional has a simpler, unified formulation from the direct model's point of view and a different method to navigate the optimality map through the Projected Newton method in the AM algorithm. Also in the cited article there is no discussion about positivity, smoothness and causality of the estimated residence time.

In the case of the~\cite{Skaggs1998} article, the direct model is similar to ours with some differences in its formulation:
\begin{equation}\label{10}
\begin{aligned}
& (\hat{f},  \hat{\alpha}) = \underset{f\in\mathbbm{R}_+^K,\alpha}{\text{argmin}} =  \frac{1}{2} \|c - A \cdot f \|_2^2 + \alpha^{2} \|\nabla^{2} f\|_{2}^{2}  \\
&\text{with } f \geq 0\ ,\  a'f=1\ ,
\end{aligned}
\end{equation}
where
\begin{itemize}
    \item $c$ is the output of the system, known;
    \item $a$ is the input of the system, known;
    \item $A$ is the Toeplitz matrix of the input of the system;
    \item $f$ is the impulse response of the system, to estimate;
    \item $\alpha$ is the hyper-parameter to estimate with Fischer's Statistic method;
    \item $\nabla^{2} f$ denotes the Hessian of $f$
\end{itemize}

The hyper-parameter $\alpha$ is here squared and determined with Fischer's Statistic method ($F$), while smoothness is implemented by a second derivative applied on $f$. There is a constraint for positivity and the condition that the integral of the obtained curve sums up to 1. The solutions are evaluated with~\cite{Provencher1982a} Fischer's Statistic method and visual inspection. Another aspect here is the multiple peak problem, where~\cite{Provencher1982a} argues to investigate separately for certain values of $F$. Also, to avoid computational difficulties in the test runs, a basis function representation of $f$ was introduced to ensure linearity between the probability density function (pdf) representation and the transport model. A causality constraint is not discussed here. In contrast, we estimate the $\alpha$ hyper-parameter ($\lambda$ in our case) by using the $SNR$ values between the reconstructed breakthrough curve and the original one. A bigger $SNR$ means a better reconstruction and also a better estimation of $k$ through the constraints, and this is realized through the $\lambda$ hyper-parameter possible choice strategies ($\alpha$ equivalent). A hydrologist can then estimate the same curve with a range of values for $\lambda$, for multiple time series and time series lengths, and then see what $\lambda$ value best fits for that particular tested site. We do smoothness regularization with a first-order derivative since testing with a second-order derivative does not show any improvement on the estimate, thus our direct model is slightly simpler. Our algorithm does not make an \textit{a priori} assumption about the shape of the estimated residence time, therefore multiple lobes can appear without having to set any fixed number of these beforehand. The estimation of $f$ ($k$ in our case) is also free of being modeled with basis functions. The sole observation here is that the channel needs to be short enough so that it can be considered linear.

In the case of~\cite{Fienen2006} the presented method is a Bayesian Monte-Carlo non-parametric deconvolution method that gives as result the full shape of the residence time distribution curve containing all possible residence time curves for that channel with zones of interest curves and the average curve. The method can yield multiple peaks in the transfer function with some computational cost -- "\textit{Using the MCMC Gibbs sampler with reflected Brownian motion requires some computational effort (CPU time up to several days on a typical desktop computer)}"~\cite{Fienen2006}. There is a constraint for positivity and for causality through~\cite{Michalak2003}. Expectation Maximization is used to estimate the parameters. The algorithm is tested on uni-modal and bi-modal cases. In comparison, our method provides faster estimates of the residence time curve for a Dirac-like rainfall event or for a clustered rainfall event. The computational cost per tested hyper-parameter $\lambda$ is small. There is no constraint on the shape of the residence time curve other than smoothness (controlled by $\lambda$), and positivity and causality which we implement throughout the algorithm. On the downside, our algorithm does not estimate the uncertainties attached to the residence time like in a Bayesian approach.

Another example is~\cite{Dietrich1993} with an algorithm based on ridge regression, where the direct model is similar to ours but has two hyper-parameters to be set.~\cite{Michalak2003} is another article where Bayesian Monte-Carlo deconvolution is done through an inverse problem setup. Here positivity and causality are implicitly enforced by the method of images applied to reflected Brownian motion that gives "\textit{a prior pdf that is non-zero only in the non-negative parameter range}"~\cite{Michalak2003}. The MCMC is here implemented with the Gibbs sampling algorithm. Similar to~\cite{Fienen2006} the result is also a pdf with zones of interest for the residence time curve. Even if the computational time for Bayesian MCMC deconvolution methods is deemed "\textit{manageable}"~\cite{Michalak2003}, probably even more so with current hardware, the need for a fast method seems necessary for the community, and we expand on this in the next paragraph.

\subsection{Comparison to the Cross-Correlation Method}

We use the cross-correlation method as a benchmark to compare the performance of our algorithm. The cross-correlation measures the similarity between two signals, the second one being a shifted version of itself.

The AM algorithm also estimates the basin measurements constant level, $\textbf{c}_{est}$, and the estimated residence time amplitude depends on this constant level. It is necessary to obtain this same amplitude for the cross-correlation method, for comparison purposes, and this is done through the following:
\begin{equation}\label{12}
\begin{aligned}
&\textbf{y}_{rec} = x \ast R_{xy}\\
&\textbf{k}_{est} = R_{xy}\cdot \dfrac{\sigma_{\textbf{y}}}{\sigma_{\textbf{y}_{rec}}}
\end{aligned}
\end{equation}
We call the cross-correlation method XCORR in our plots.

The cross-correlation implicitly assumes that the input rainfall is white noise. In this case, the auto-correlation of each rain fall time series would be a Dirac at the center. Since real rainfall time series have actually long-tailed statistics, the cross-correlation method is inexact. Here we use multifractals to simulate realistic rainfall~\cite{Z_Hydro_Tessier1996b}. Therefore, we expect the cross-correlation method to have a limited performance in real life tests.

The decision to benchmark against the cross-correlation is due to the fact that it is the preferred method for hydrologists in numerous recent articles: for determining transport of biological constituents in~\cite{Sheets2002}, or studying river-groundwater interaction with different types of measurements being cross-correlated like in~\cite{HoehnCirpka2006}. Cross-correlation is also used by~\cite{Vogt2010} for estimating mixing ratios and mean residence times, by~\cite{Z_Hydro_Delbart_2014} for estimating the pure residence time curve. Therefore, the hydrology community is interested in a simple and fast method with minimal implementation time that gives a residence time curve estimation from different time series measurements. In the case of the cross-correlation method, one focuses on analyzing the position of the maximal amplitude and general shape of the curve. From this curve hydrologists extract the characteristics of interest for that particular channel (mean residence time, mixing ratios, etc.). In contrast to the cross-correlation method we offer positivity, smoothness and causality constraints that give a more precise curve and a similar computing time.

\subsection{Comparison to\texorpdfstring{~\cite{Z_Hydro_Cirpka2007}}{Cirpka Method}}

Another benchmark method for the AM is the one presented in~\cite{Z_Hydro_Cirpka2007} that uses measurements in fluctuations of electrical-conductivity as inputs, with a direct model similar to~\eqref{1}.
The algorithm in~\cite{Z_Hydro_Cirpka2007} is the same as the one used in~\cite{Vogt2010} and both articles compare their results with those of the cross-correlation method. In~\cite{Z_Hydro_Cirpka2007} the deconvolution algorithm is also an Alternating Minimization algorithm, but this time between estimating the residence time in the first step using a Bayesian Maximum A Posteriori method, and estimating the variance of the noise and the slope parameters in the second step. One can notice that Equation~\eqref{3} is similar to~\cite[Eq.(8)]{Z_Hydro_Cirpka2007}.
One main advantage of the~\cite{Z_Hydro_Cirpka2007} approach is that it delivers the uncertainty curves of the full Bayesian method while not being a full Bayesian deconvolution method, thus having a fast computation time. One drawback is that the two parameters, variance of noise and slope, need to have well chosen initial values. In a full Bayesian based deconvolution these parameters would also need to be estimated and this would be done by Markov Chain Monte Carlo methods which are computationally intensive. With regularization based deconvolution we try to avoid high computational costs and having multiple parameters that need carefully chosen initial values. The optimal value for our hyper-parameter $\lambda$ can be automatically obtained from the inputs.

\section{Synthetic Data}
\label{sec:synthetic}

\subsection{General Discussion and \texorpdfstring{$\lambda$}{lambda} Choice Strategies}

In the context of a realistic synthetic validation we generate the rain signals $\textbf{x}$ with a multifractal simulation based on~\cite{Z_Hydro_Tessier1996b}. We use the multifractal parameters $H=-0.1,C1=0.4,\alpha=0.7$.
Furthermore we simulate  $\textbf{k}$  with a Beta distribution $B(x,\alpha=2,\beta=6)$.
We choose arbitrarily $c =  100$. To evaluate the computed estimates we use the SNR definition, where we replace the noise term with the estimated $\textbf{k}_{est}$ signal or the $\textbf{y}_{rec}$ signal respectively.
\begin{equation}\label{11}
\begin{aligned}
&SNR = 20 \log_{10}\dfrac{\|m\|_{2}^{2}}{\|m - m_{est}\|_{2}^{2} }[dB]\ ,
\end{aligned}
\end{equation}
where $m$ is the true signal $\textbf{k}$ or $\textbf{y}$ and $m_{est}$ is the estimated $\textbf{k}_{est}$ or reconstructed $\textbf{y}_{rec}$ signal respectively.

\begin{figure}[H]
    \centering
    \subfigure[]{\label{subfig_6}\includegraphics[clip,trim=2.3cm 1.7cm 2.7cm 1.8cm,scale=0.4]{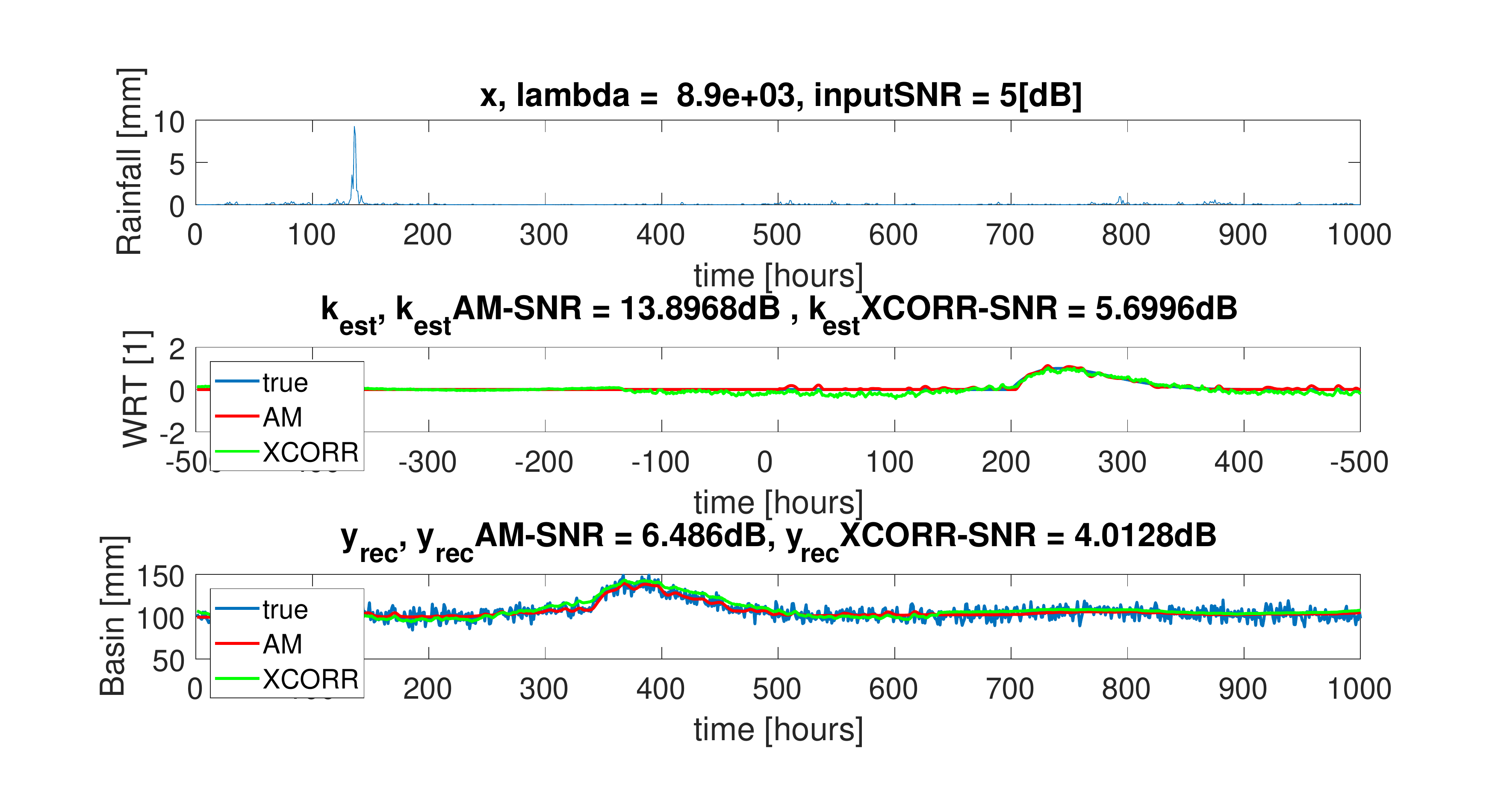}}
    \subfigure[]{\label{subfig_7}\includegraphics[clip,trim=2.3cm 1.7cm 2.7cm 1.8cm,scale=0.4]{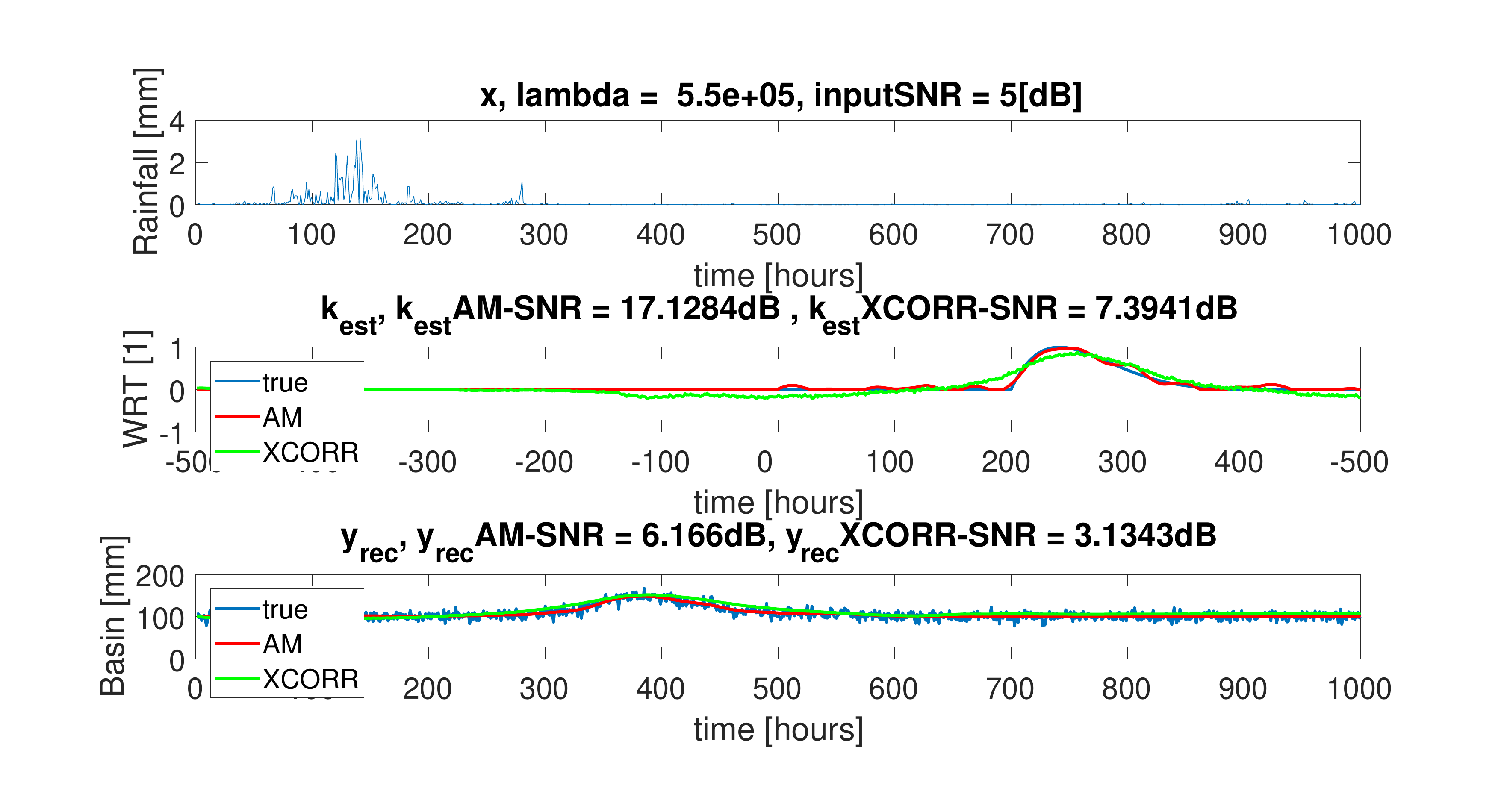}}
    \caption{Two examples of the residence time estimation $\textbf{k}_{est}$ and reconstructed aquifer levels $\textbf{y}_{rec}$ from synthetic data for a $\textbf{y}$ input SNR of 5 dB (noisy measurements). The input rain is generated with realistic multifractal time series. AM stands for the Alternating Minimization, XCORR for the standard cross-correlation, true for the true solution.}
    \label{fig_synthetic_1}
\end{figure}

\begin{figure}[H]
    \centering
    \subfigure[]{\label{subfig_8}\includegraphics[clip,trim=2.3cm 1.7cm 2.7cm 1.8cm,scale=0.4]{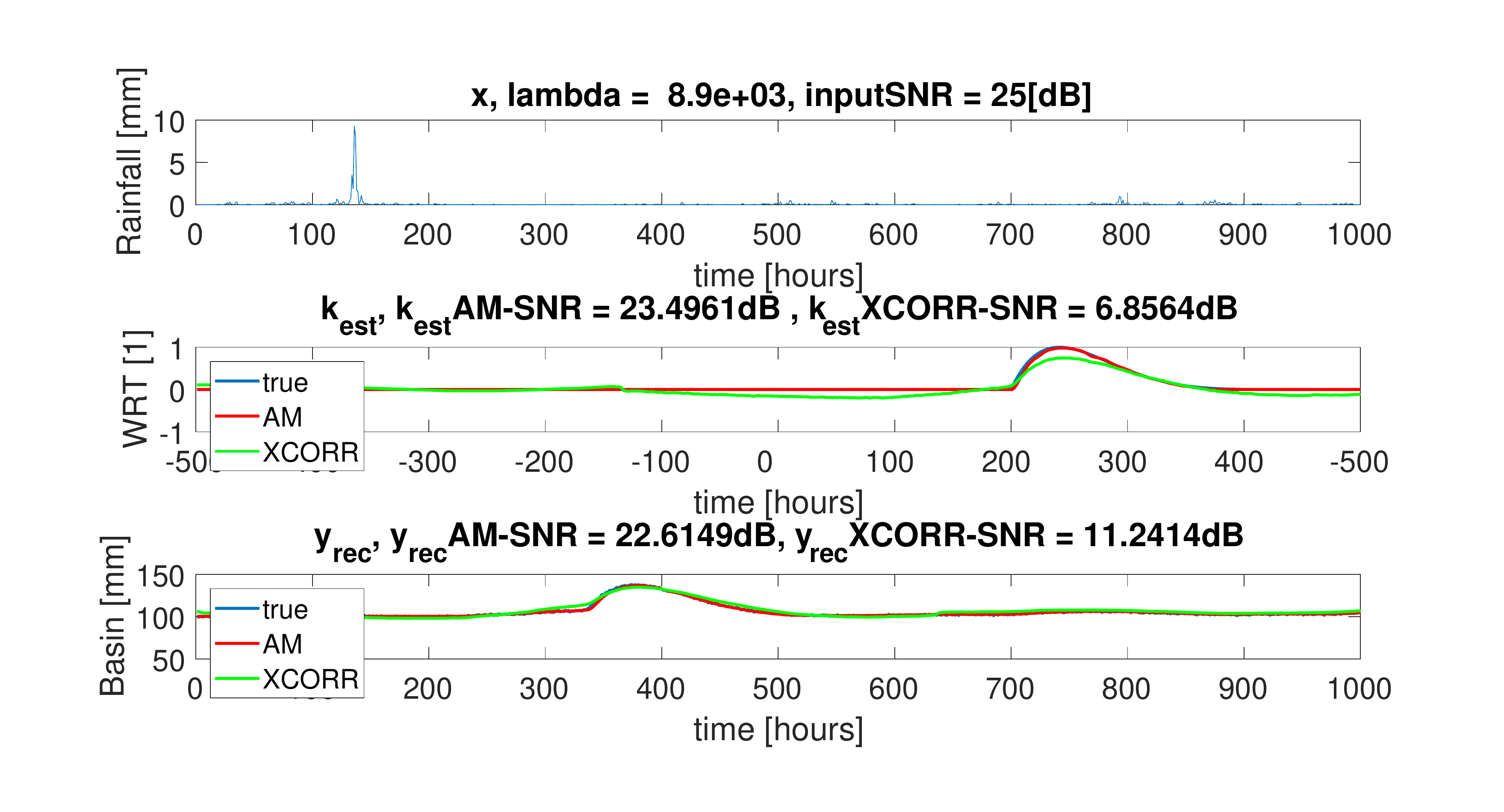}}
    \subfigure[]{\label{subfig_9}\includegraphics[clip,trim=2.3cm 1.7cm 2.7cm 1.8cm,scale=0.4]{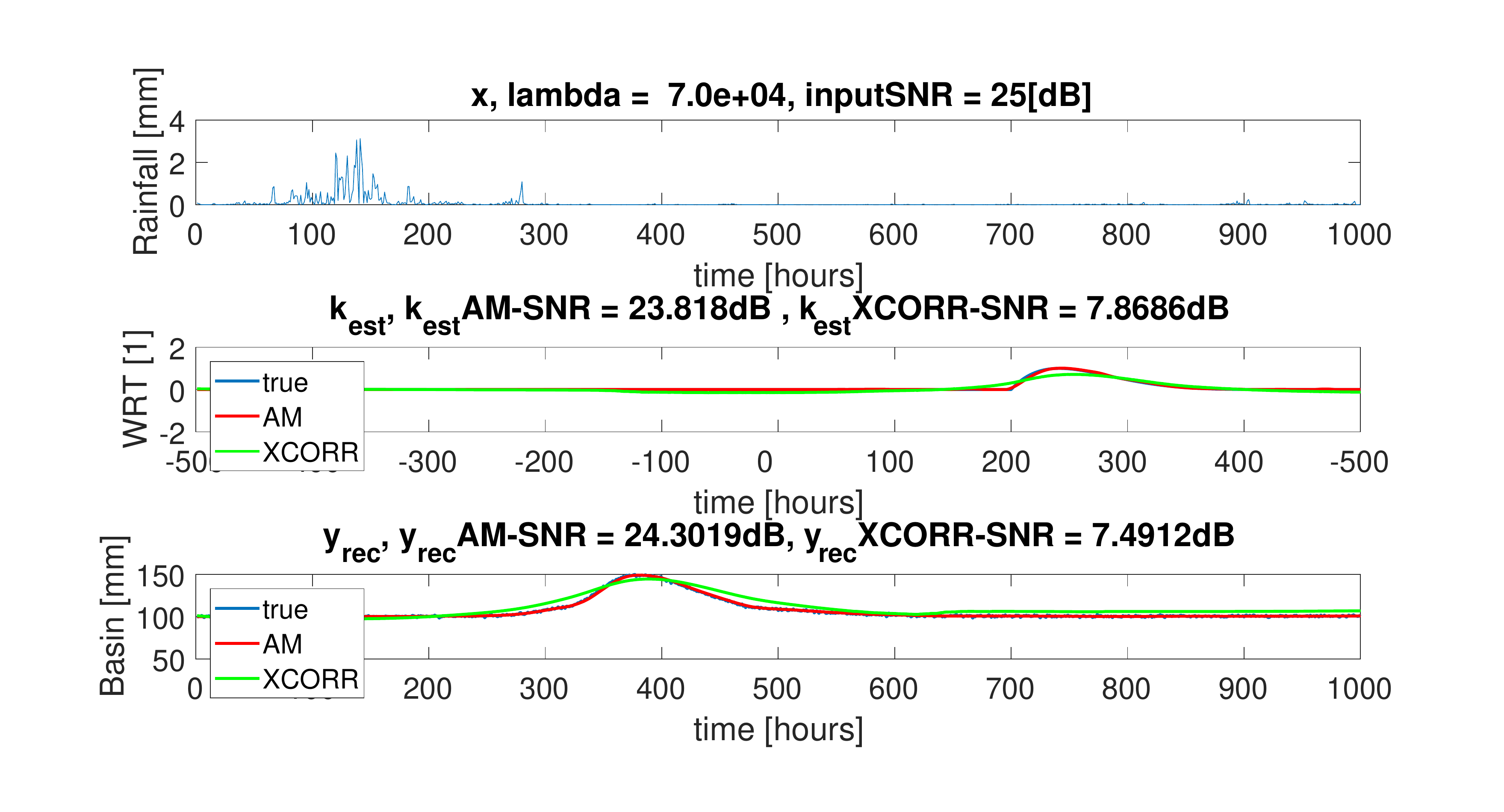}}
    \caption{Same as in Figure ~\ref{fig_synthetic_1} for a $\textbf{y}$ input SNR of 25 dB.}
    \label{fig_synthetic_2}
\end{figure}

Examples of results obtained from synthetic data are shown in Figure~\ref{fig_synthetic_1} and Figure~\ref{fig_synthetic_2}.
The positivity and causality constraints are well respected. In addition, our method always provides a better estimation of the residence time $\textbf{k}_{est}$ in comparison with the standard cross-correlation method. The cross-correlation method manages to preserve the position of the maximum intensity of the residence time distribution but does not match either the shape or the amplitude of the true $\textbf{k}$.
It can be observed that for a high noise level of $\textbf{y}$, the $\lambda$ hyper-parameter must be greater in order to obtain better estimates $\textbf{k}_{est}$ and $\textbf{y}_{rec}$. The greater the $\lambda$, the greater the importance of the regularization term in comparison to the fidelity term therefore smoothing is more important, which improves results when entries are noisy. Therefore, an analysis of the deconvolution results is also necessary in order to find the right adaptation of the $\lambda$ hyper-parameter for a particular noise level. 

We propose four strategies to automatically tune the $\lambda$ hyper-parameter.
\begin{enumerate}
    \item $\lambda_{oracle}$: choosing the $\lambda$ corresponding to the best estimation of $\textbf{k}_{est}$ by maximizing the $\textbf{k}_{est}$ SNR output (or minimizing the distance between $\textbf{k}_{est}$ and $\textbf{k}$). This strategy only works if the solution is known and represents the maximum achievable value.
    \item $\lambda_{discrepancy}$: choosing the $\lambda$ giving the residual variance between $\textbf{y}$ and $\textbf{y}_{rec}$ closest to that of the noise. This method is known as "Morozov's discrepancy principle" ~\cite{Z_Hydro_Pereverzev2009}.
    \item $\lambda_{fidelity}$: choosing the $\lambda$ corresponding to the best reconstruction of $\textbf {y}_{rec}$ by maximizing the $\textbf{y}_{rec}$ SNR output (or minimizing the distance between $\textbf{y}_{rec}$ and $\textbf{y}$). This is the value of the reconstruction optimum. This completely heuristic method automatically selects the hyper-parameter with a performance close to the selection by "discrepancy principle" as will be seen next, in a completely blind way (without \textit{a priori} knowledge of the variance of the noise).
    \item $\lambda_{corrCoeff}$: choosing the $\lambda$ corresponding to the best reconstruction of $\textbf {y}_{rec}$ by maximizing the correlation coefficient value between $\textbf{y}_{rec}$ and $\textbf{y}$.
\end{enumerate}

The four $\lambda$ strategies give different estimates of $\textbf{k}_{est}$, whose SNR value is compared to the $\textbf{y}$ input SNR (measurements noise level), the goal being to obtain the best possible $\textbf{k}_{est}$ SNR for each given $\textbf{y}$ input SNR level. The algorithm is tested for different input SNR values from 0 dB (very high noise level) to 30 dB (almost no noise) and over a $\lambda$ range chosen from $10^{-5}$ to $10^{12}$  with 20 values dispersed on a logarithmic scale.

To show the quality of estimation, for each noise level, we run arbitrarily 30 test cases (input rainfall $\textbf{x}$). For each randomly chosen $\textbf{x}$ convolved with the known $\textbf{k}$, the resulting $\textbf{y}$ signal has Gaussian noise added to it according to the input SNR test value. 
We apply the AM, XCORR and \cite{Z_Hydro_Cirpka2007} methods to each test case for all $\lambda$s.
For each test run we record the $\textbf{k}_{est}$ SNR value, the $\textbf{y}_{rec}$ SNR value and the $\textbf{y}_{rec}$ correlation coefficient. Since 30 tests are made for each input $\textbf{y}$ SNR, we obtain 30 plots showing the evolution of the $\textbf{k}_{est}$ SNR, of $\textbf{y}_{rec}$ SNR and $\textbf{y}_{rec}$ correlation coefficient, depending on the $\lambda$ choice. 

\begin{figure}[H]
    \centering
    \subfigure[]{\label{subfig_10}\includegraphics[clip,trim=2.3cm 1.7cm 2.7cm 1cm,scale=0.25]{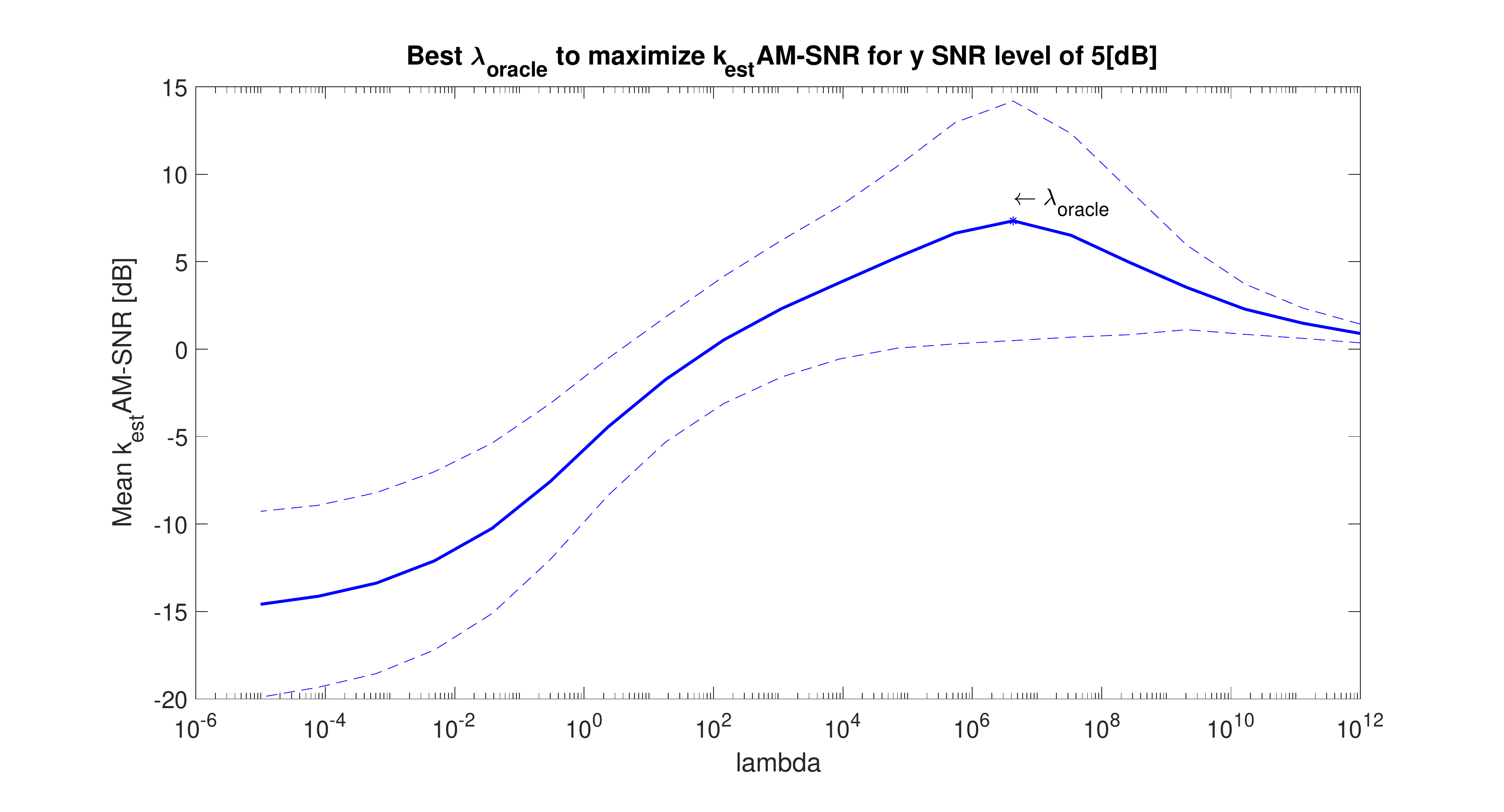}}
    \subfigure[]{\label{subfig_11}\includegraphics[clip,trim=2.3cm 1.7cm 2.7cm 1cm,scale=0.25]{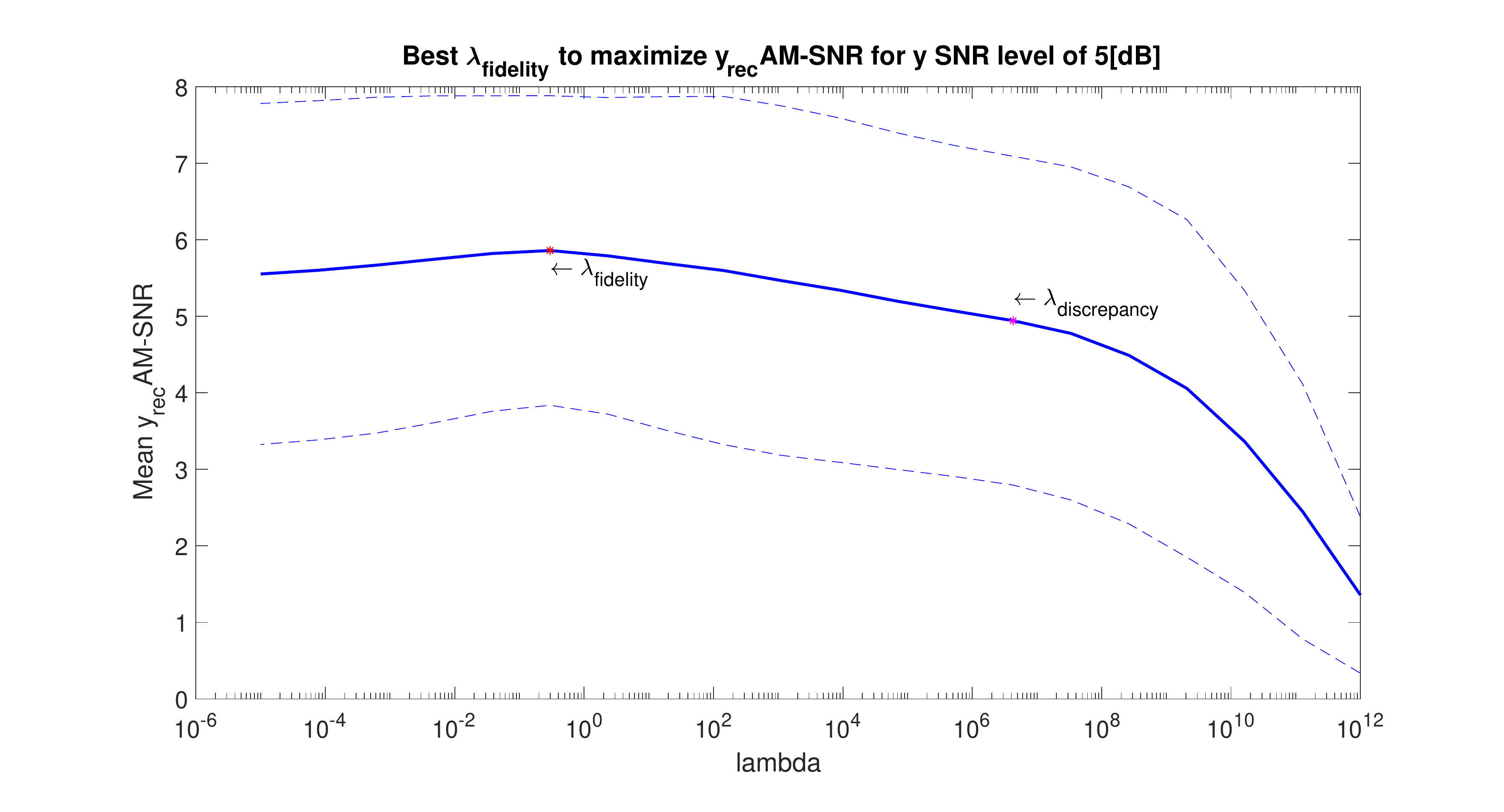}}
    \subfigure[]{\label{subfig_12}\includegraphics[clip,trim=2.3cm 1.7cm 2.7cm 1cm,scale=0.25]{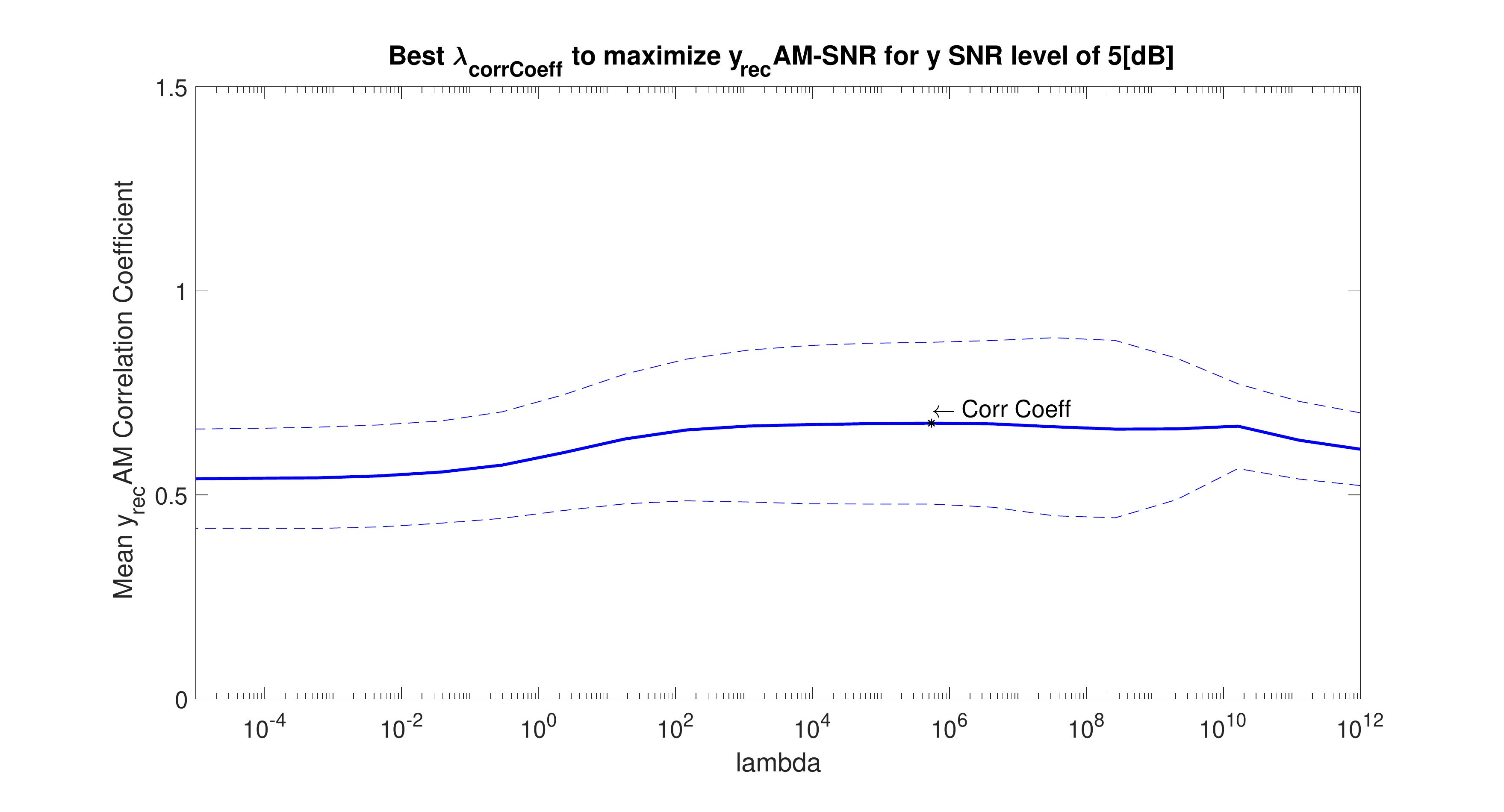}}
    \caption{Selection strategy of hyper-parameter $\lambda$. We plot average and standard deviation over 30 synthetic examples of: (a) $\textbf{k}_{est}$ SNR, (b) $\textbf{y}_{rec}$ SNR and (c) $\textbf{y}_{rec}$ correlation coefficient as a function of $\lambda$.  The $y$ input SNR is 5 dB, meaning very noisy measurements. The $\lambda_{oracle}$ point in (a) shows the best $\lambda$ in average to maximize the $\textbf{k}_{est}$ SNR for the synthetic tests. This can be computed only when the true solution is known. In (b) the $\lambda_{fidelity}$ maximizes the $\textbf{y}_{rec}$ SNR. The $\lambda_{discrepancy}$ is achieved when $\textbf{y}_{rec}$ SNR is closest to the actual noise level. In (c), the $\lambda_{corrCoeff}$ is the optimum over the correlation coefficient between $\textbf{y}_{rec}$ and $\textbf{y}$.}
    \label{fig_SNR_lambdas_5}
\end{figure}

\begin{figure}[H]
    \centering
    \subfigure[]{\label{subfig_13}\includegraphics[clip,trim=2.3cm 1.7cm 2.7cm 1cm,scale=0.25]{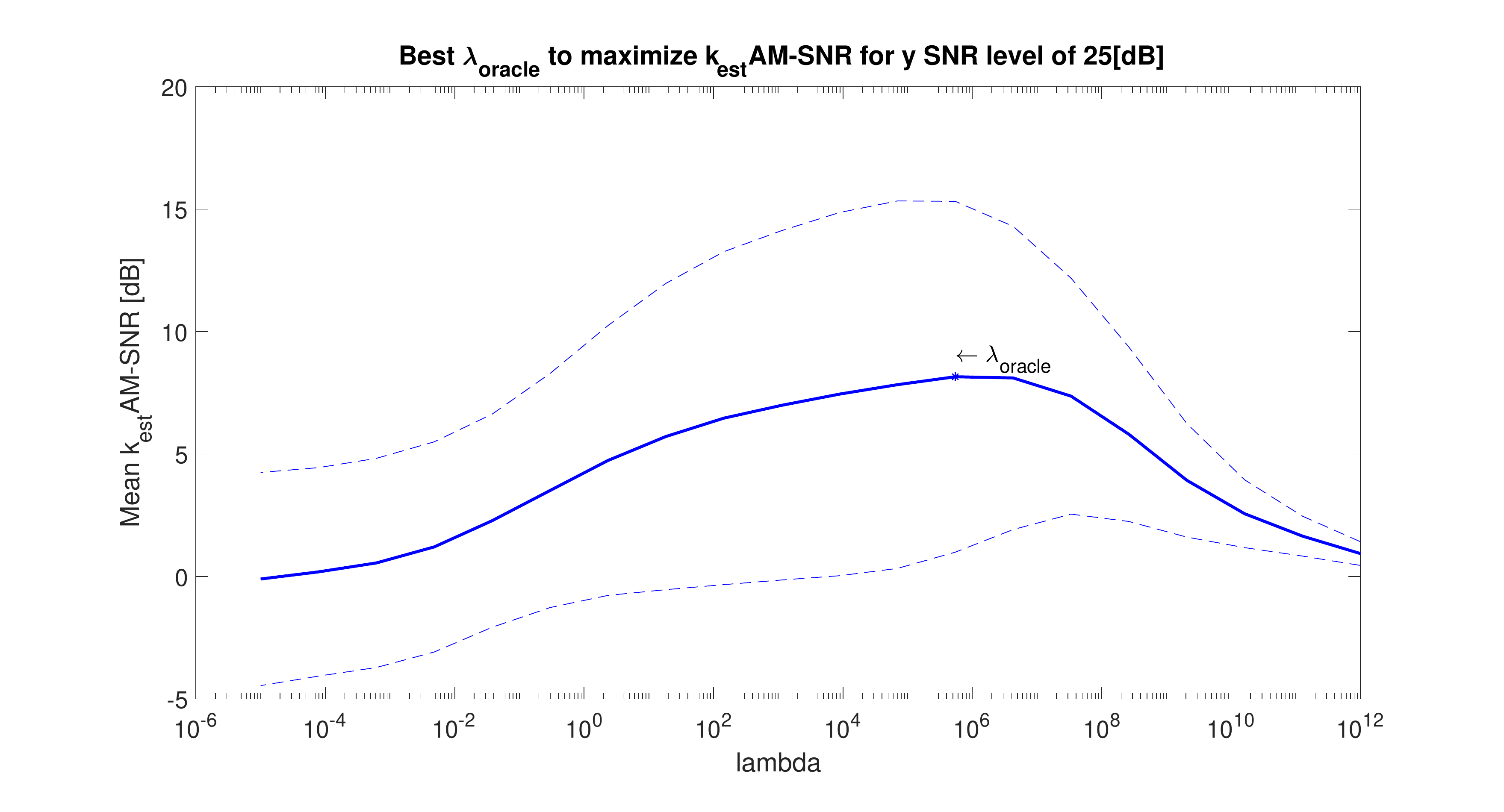}}
    \subfigure[]{\label{subfig_14}\includegraphics[clip,trim=2.3cm 1.7cm 2.7cm 1cm,scale=0.25]{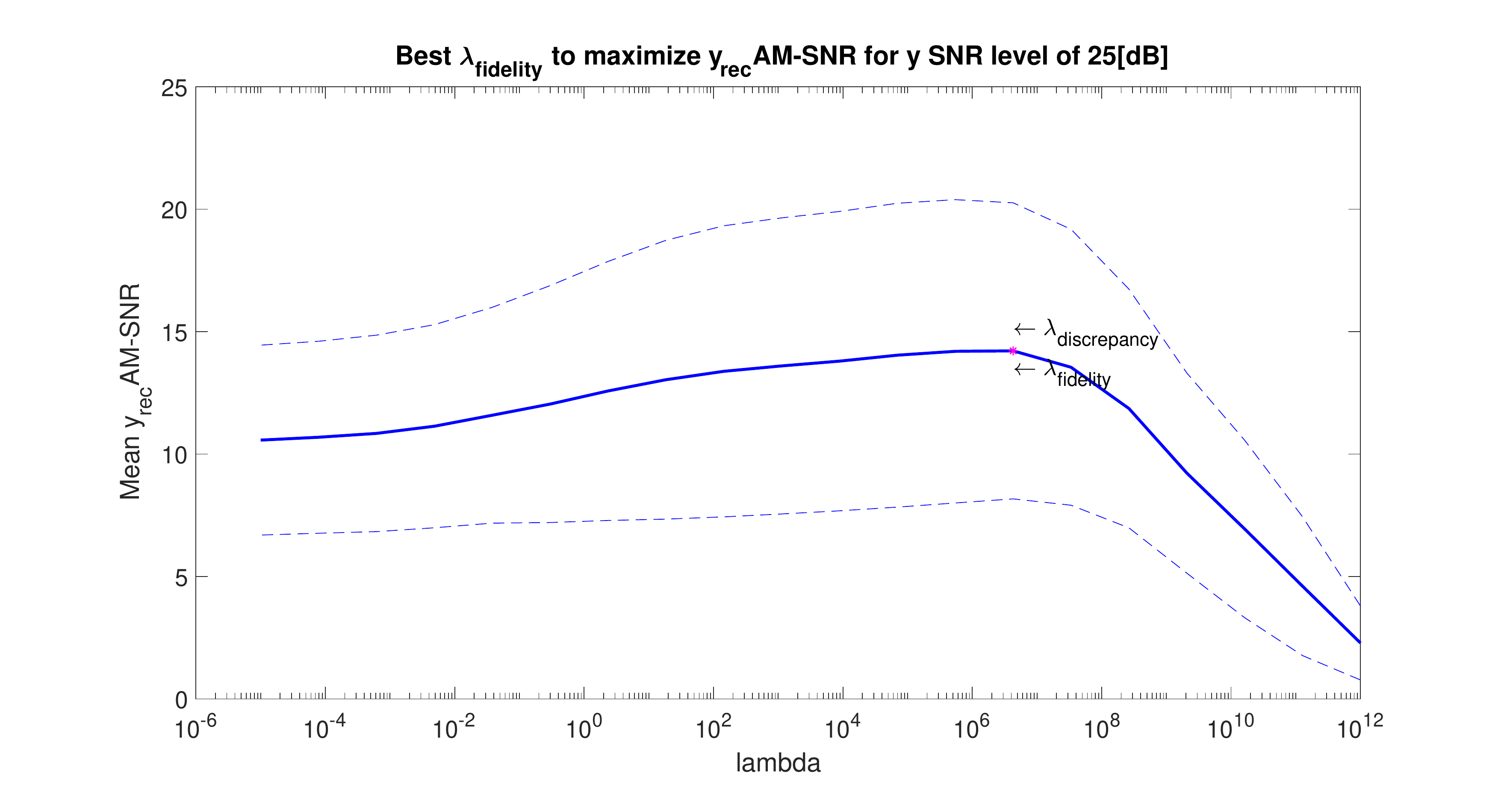}}
    \subfigure[]{\label{subfig_15}\includegraphics[clip,trim=2.3cm 1.7cm 2.7cm 1cm,scale=0.25]{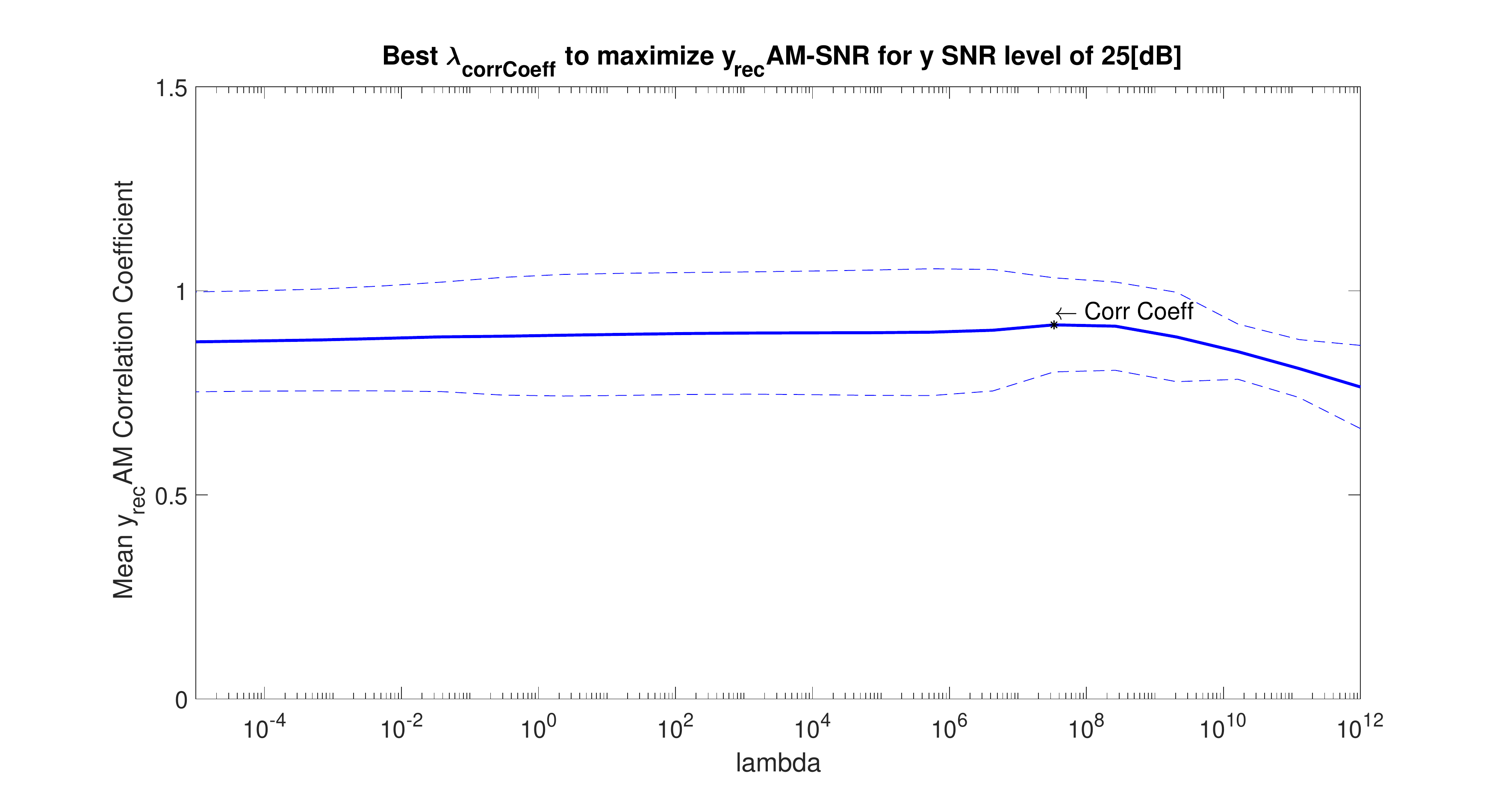}}
    \caption{Same as in Figure ~\ref{fig_SNR_lambdas_5} with an $\textbf{y}$ input SNR of 25 dB. We find that $\lambda_{fidelity}$, $\lambda_{discrepancy}$ and $\lambda_{corrCoeff}$ approach the optimal $\lambda_{oracle}$ in average.}
    \label{fig_SNR_lambdas_25}
\end{figure}

By averaging these plots, mean values and their standard deviation can be computed which are shown in Figure~\ref{fig_SNR_lambdas_5} for a $\textbf{y}$ input SNR of 5 dB and Figure~\ref{fig_SNR_lambdas_25} for 25 dB respectively. We lose the optimality for each single example due to averaging, but we show the variability of the criteria depending on noise level and input data. We also present graphically the four strategies of $\lambda$ determination.

\begin{figure}[H]
    \centering
    \subfigure[]{\label{subfig_16} \includegraphics[trim=1cm 2cm 1cm 2cm, scale=0.32]{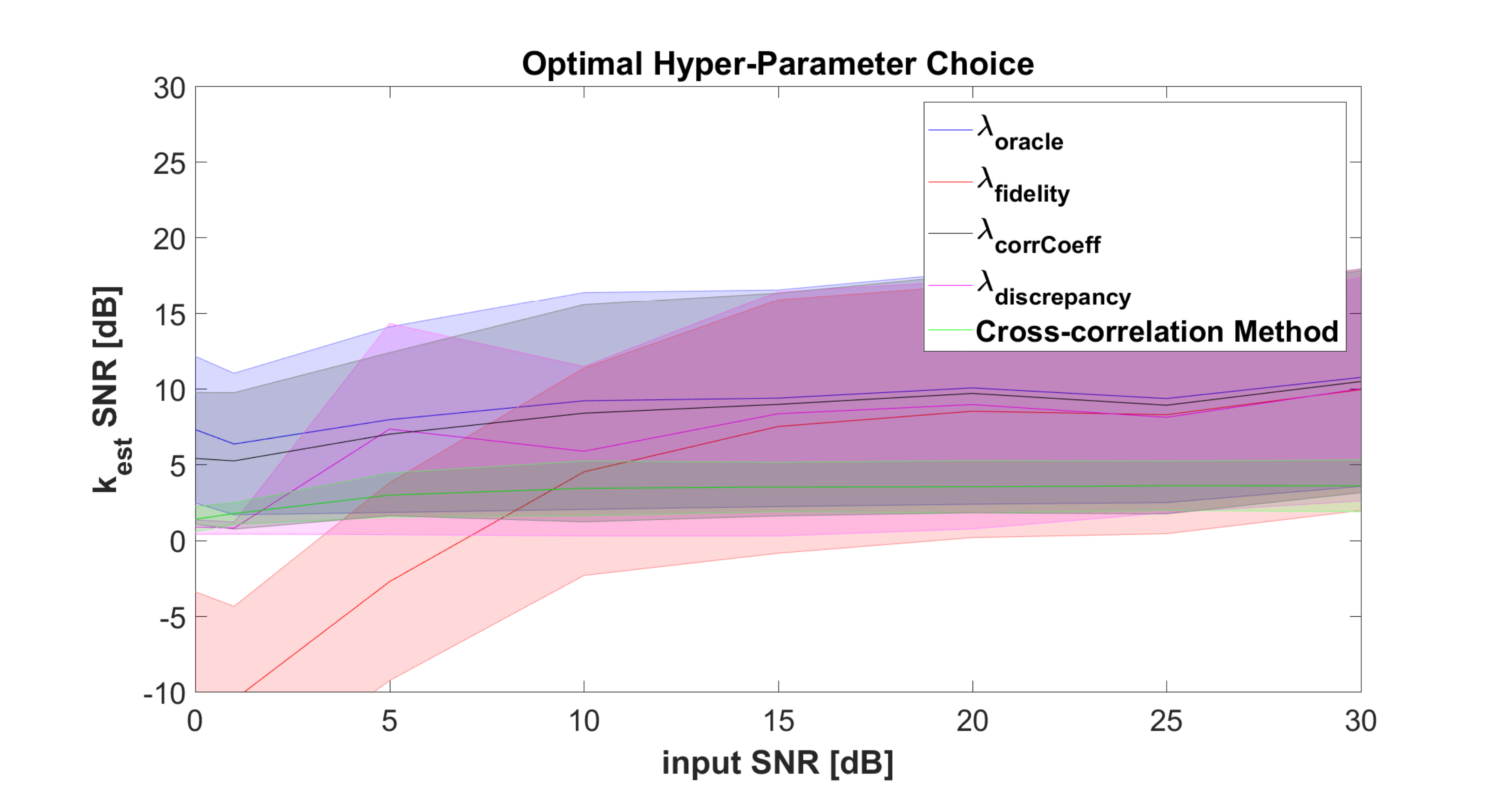}}
    \subfigure[]{\label{subfig_17} \includegraphics[trim=1cm 2cm 1cm 2cm, scale=0.32]{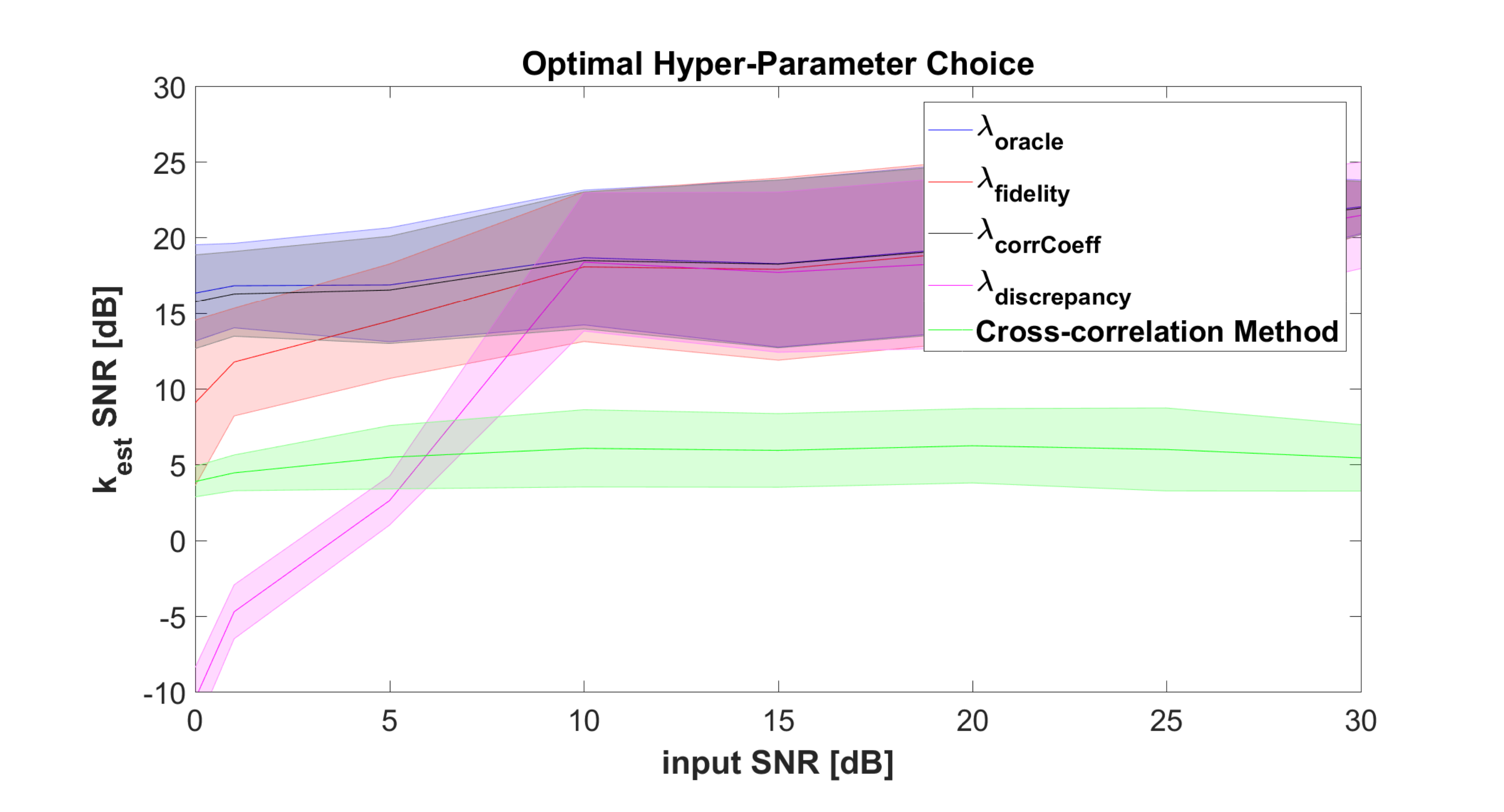}}
    \caption{Quality of the residence time estimation $\textbf{k}_{est}$ for the four hyper-parameter selection strategies and the cross-correlation method. Mean and standard deviation of obtained $\textbf{k}_{est}$ SNRs, as a function of the noise level of the measurements, for inputs of length: 1000 data points (a) and 5000 data points (b). The cross-correlation method always stands lower indicating a poorer estimation. The correlation coefficient strategy $\lambda_{corrCoeff}$ is the best strategy, across noise level and signal length.}
    \label{fig_hyper_parameter_choice_strategy}
\end{figure}

In Figure~\ref{fig_hyper_parameter_choice_strategy}, we can see how the four strategies compare with the cross-correlation method. For a $\textbf{k}_{est}$ length of 1000 data points to estimate, we show in (a) the results for when inputs $\textbf{x}$ and $\textbf{y}$ are 1000 data points long and in (b) the results for when they are 5000 data points long. The $\textbf{k}_{est} SNR$ is always the best for the $\lambda_{oracle}$ strategy as expected. Across the plots, $\lambda_{corrCoeff}$ performs closest to it. The $\lambda_{fidelity}$ strategy is similar to $\lambda_{discrepancy}$ for SNRs from $10$ dB to $30$ dB. For the highest noise level, $\textbf{y}$ input $SNR<$ 10 dB, $\lambda_{fidelity}$ is worst for short time series and $\lambda_{discrepancy}$ is worst for longer time series. Whatever the strategy, our method is always better than the cross-correlation. 

\begin{figure}[H]
    \centering
    \subfigure[]{\label{subfig_18} \includegraphics[trim=1cm 2cm 1cm 2cm, scale=0.32]{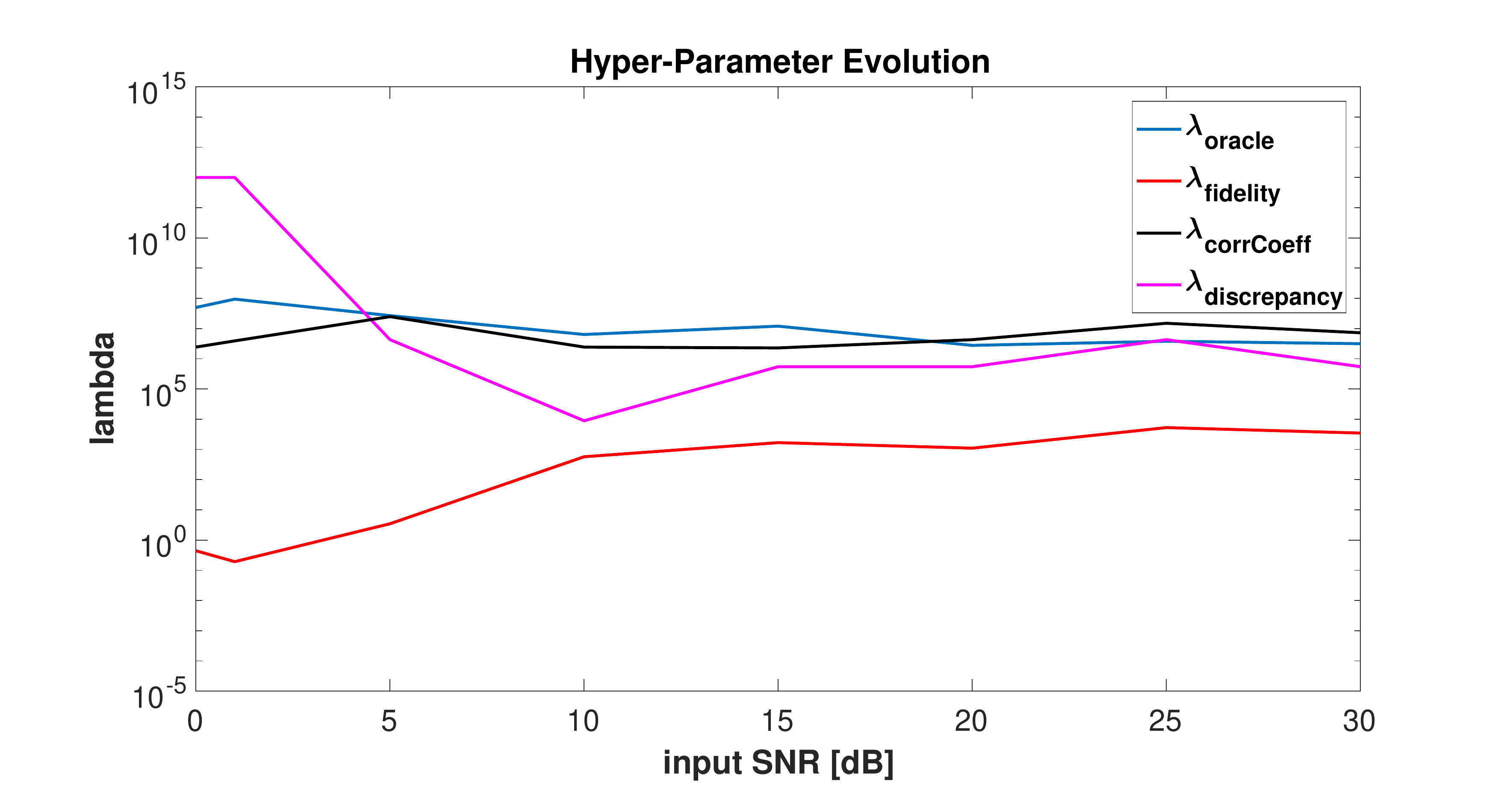}}
    \subfigure[]{\label{subfig_19} \includegraphics[trim=1cm 2cm 1cm 2cm, scale=0.32]{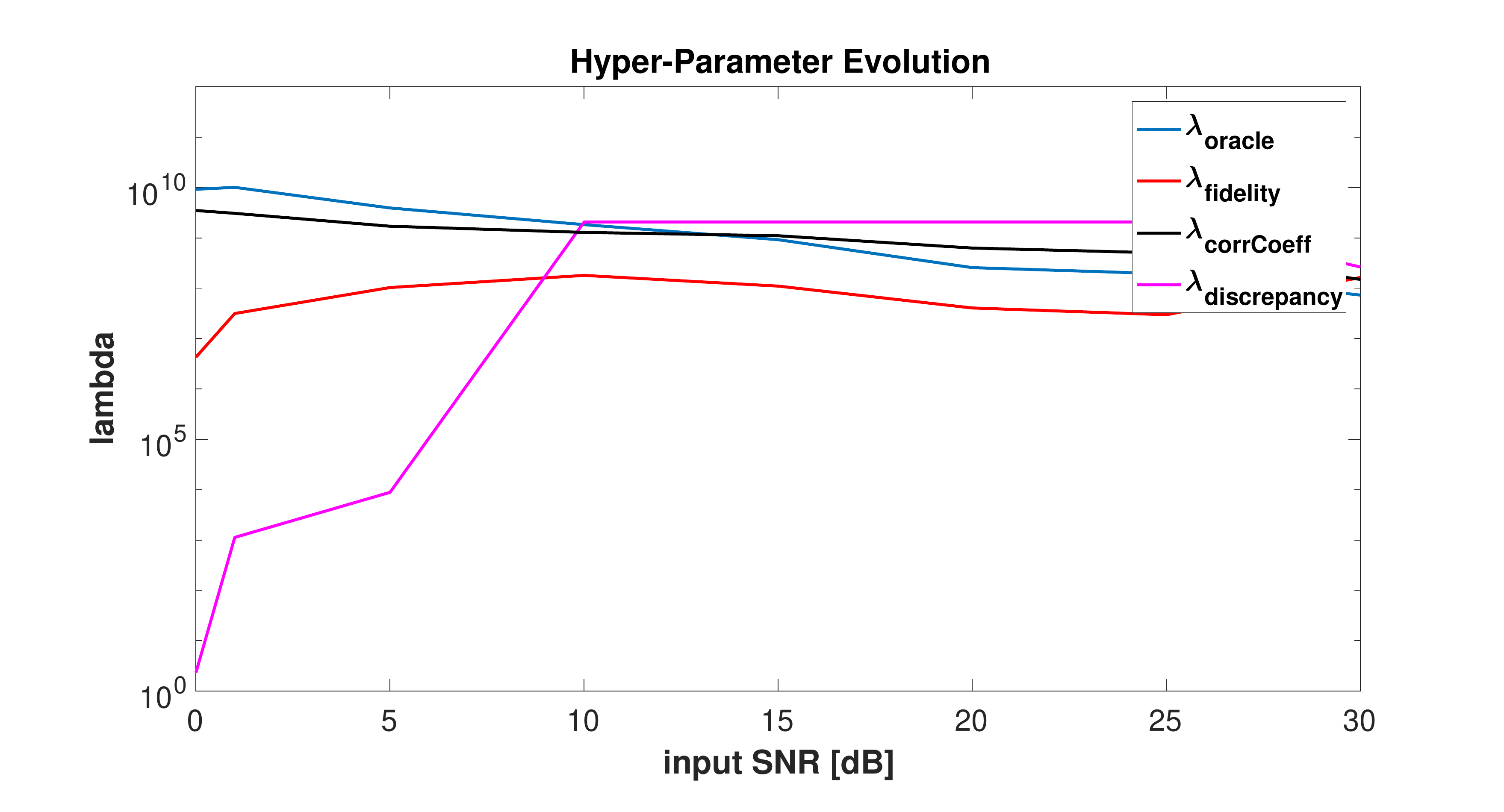}}
    \caption{The evolution of the four $\lambda$ strategies depending on the input SNR. For 1000 data points in (a) and 5000 data points in (b).}
    \label{fig_hyper_parameter_choice_strategy_evolution}
\end{figure}

The average optimal $\lambda$ value for each strategy, given the $\textbf{y}$ input SNR level, is presented in Figure~\ref{fig_hyper_parameter_choice_strategy_evolution}. In (a) and (b), we see the evolution of the $\lambda$ values versus the $\textbf{y}$ input SNR for the four given strategies. The four strategies of the hyper-parameters $\lambda$ are similar at low noise level, down to 10 dB for both 1000 and 5000 data points. Then, they begin to diverge but  $\lambda_{corrCoeff}$ always stays in the neighborhood of  $\lambda_{oracle}$, meaning it is a valid strategy to use in real test cases where $\textbf{k}$ is not known. At very high noise levels for 1000 data points,  $\lambda_{discrepancy}$ increases and provides an over-regularized, highly smooth solution that is far from the optimum. For 5000 data points both $\lambda_{fidelity}$ and $\lambda_{discrepancy}$ deliver smaller $\lambda$s. If for $\lambda_{fidelity}$ we can still expect that it would deliver a proper $\textbf{k}_{est}$, we can suspect that $\lambda_{discrepancy}$ would stress more an attachment to the data. This means that the estimated $\textbf{k}_{est}$ would give a $\textbf{y}_{rec}$ that would follow too closely the shape of $y$, including its noise.

\begin{figure}[H]
    \centering
    \subfigure[]{\label{subfig_20} \includegraphics[trim=1cm 2cm 1cm 2cm, scale=0.32]{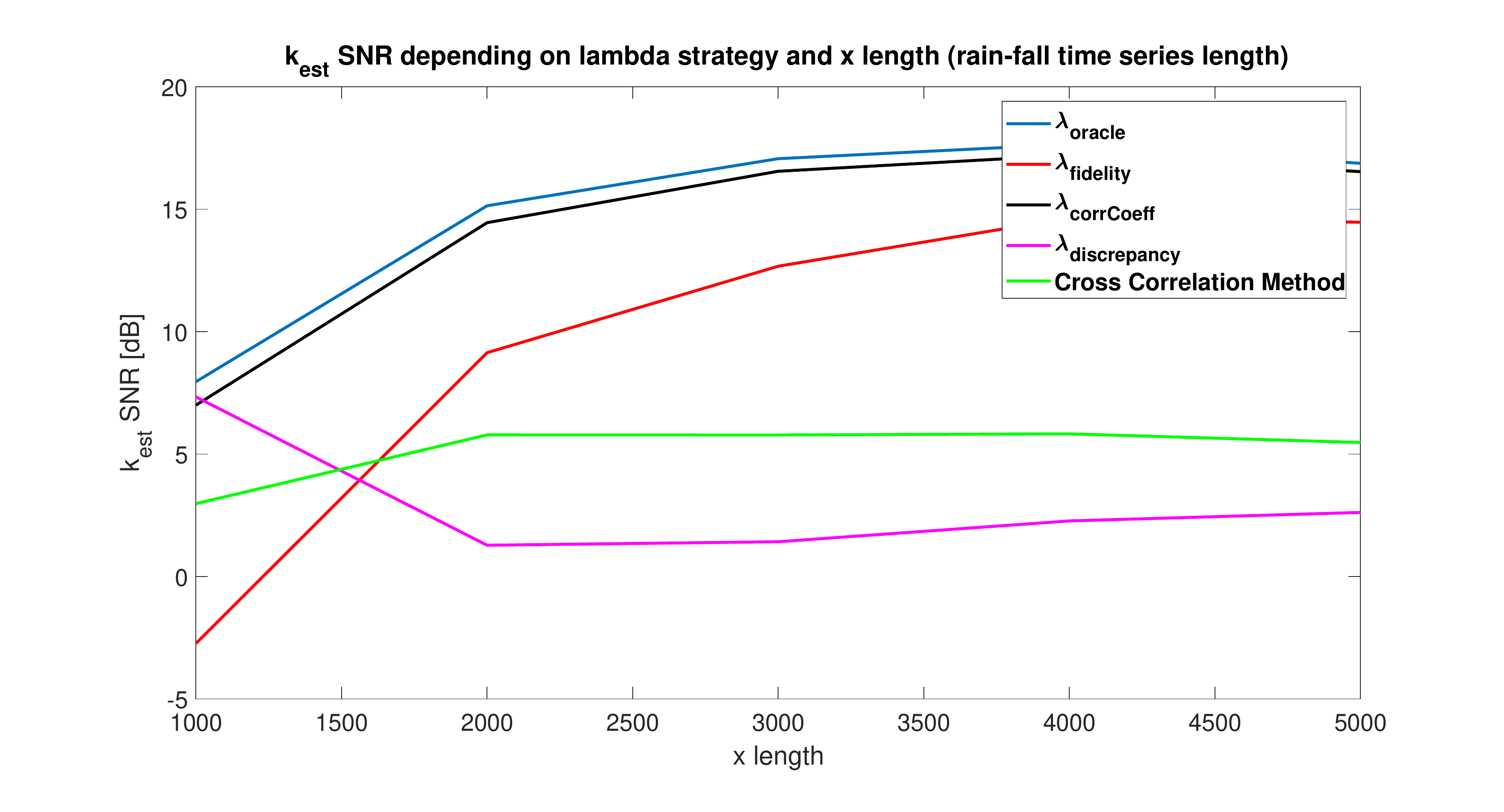}}
    \subfigure[]{\label{subfig_21} \includegraphics[trim=1cm 2cm 1cm 2cm, scale=0.32]{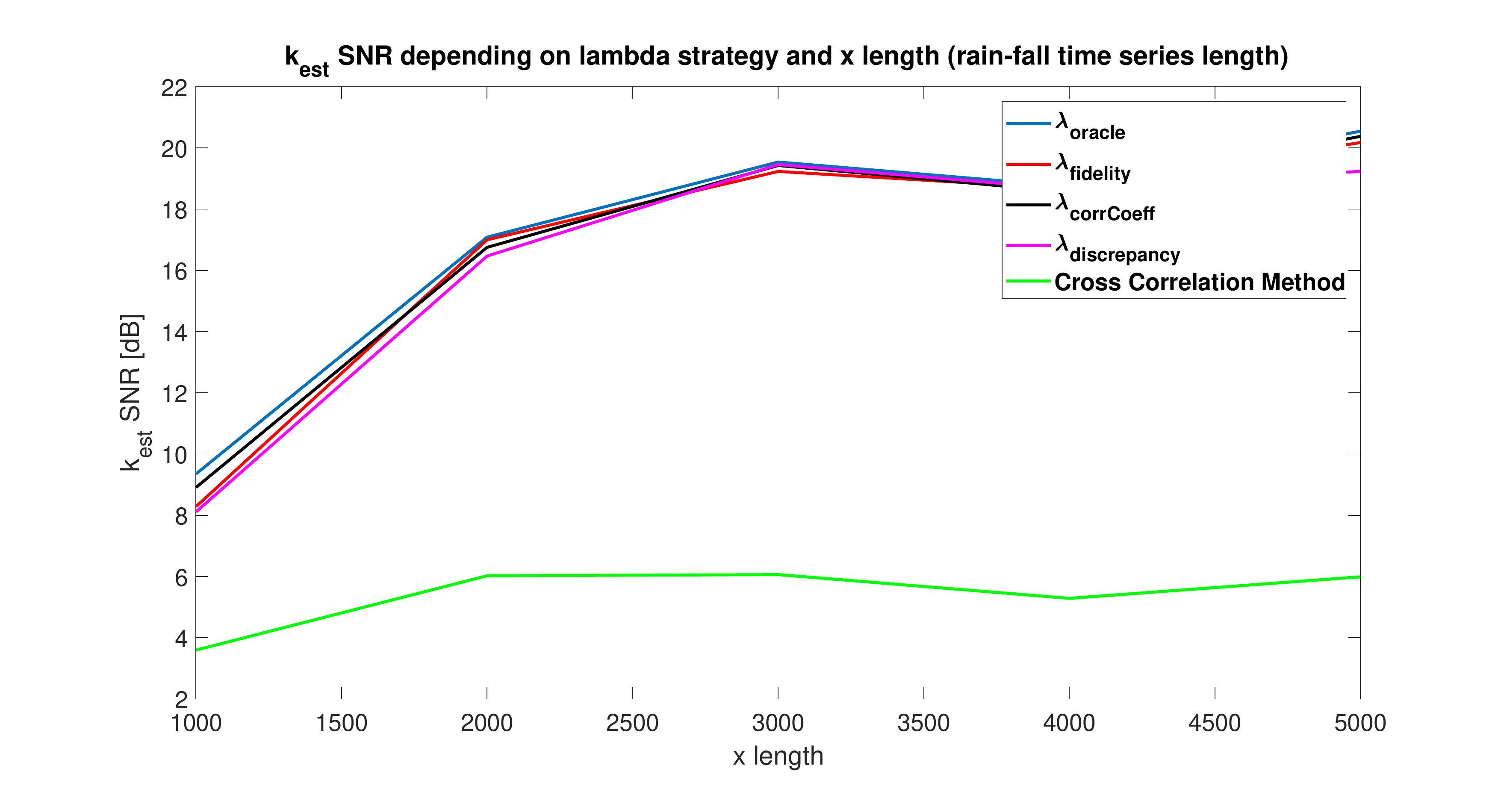}}
    \caption{Quality of residence time $\textbf{k}_{est}$  estimation depending on the number of data points contained by $\textbf{x}$ (input rain) and $\textbf{y}$ (output aquifer level). We can observe that more data points lead to a better estimation for our method for all four  $\lambda$  strategies. (a) is for a $\textbf{y}$ input SNR of 5 dB and (b) is for a $\textbf{y}$ input SNR of 25 dB}
    \label{fig_kSNR_vs_length}
\end{figure}

Furthermore we investigate the influence of data volume on the $\textbf{k}$ estimate. The aggregated results are presented in Figure~\ref{fig_kSNR_vs_length}, (a) for a $\textbf{y}$ input SNR of 5 dB and in (b) for a $\textbf{y}$ input SNR of 25 dB. All of our four strategies show significant improvement when the input time series of rainfall and aquifer measurements are longer, especially when the measurements are noisy.

\subsection{Comparison to Similar Methods}

\begin{figure}[H]
    \centering
    \subfigure[]{\label{subfig_22} \includegraphics[trim=1cm 2cm 1cm 2cm, scale=0.32]{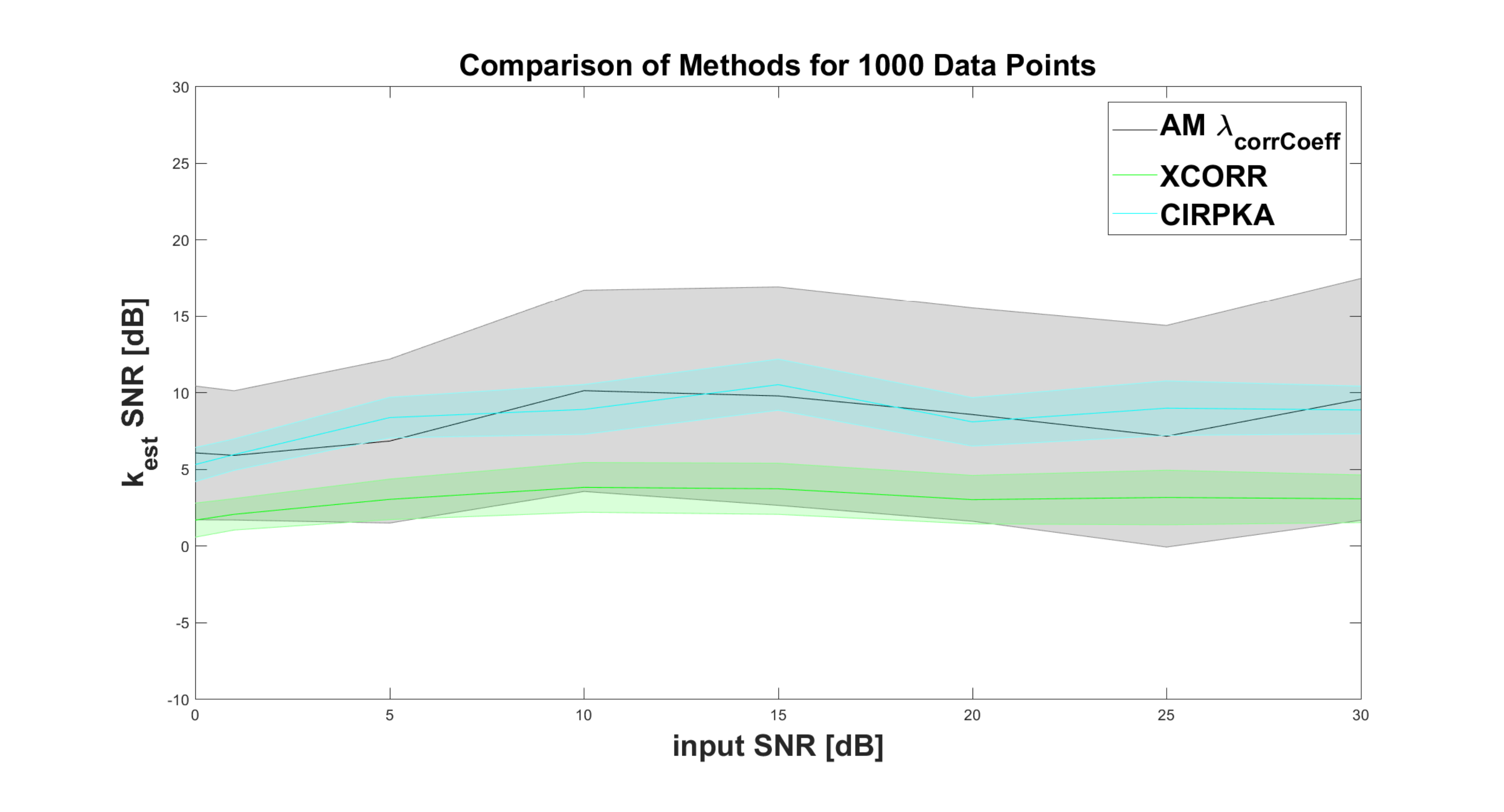}}
    \subfigure[]{\label{subfig_23} \includegraphics[trim=1cm 2cm 1cm 2cm, scale=0.32]{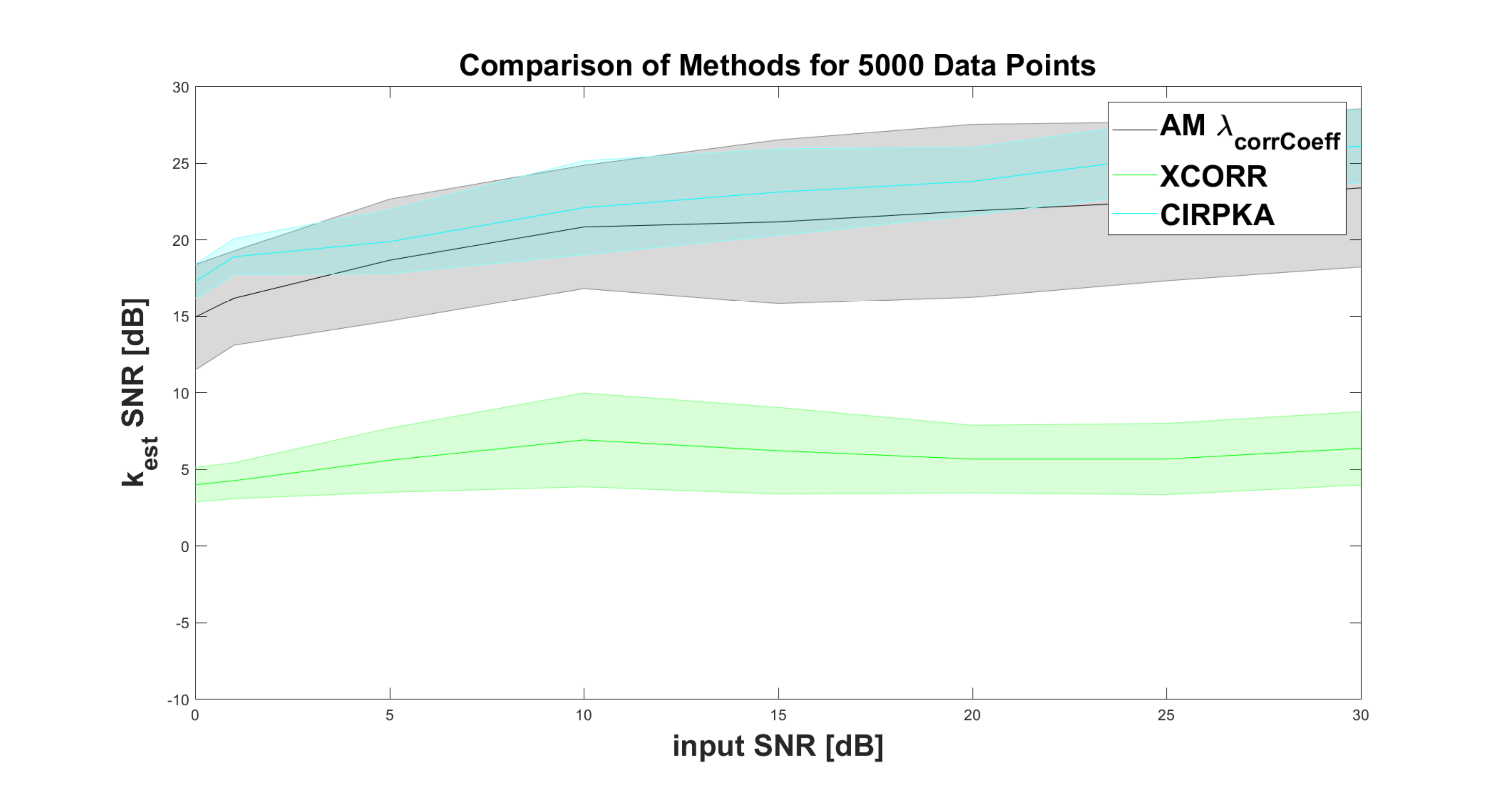}}
    \caption{Comparison between our algorithm, the cross-correlation and the~\cite{Z_Hydro_Cirpka2007} algorithm for 1000 data points (a) and 5000 data points (b)}
    \label{fig_comparison_to_xcorr_cirpka}
\end{figure}

In Figure~\ref{fig_comparison_to_xcorr_cirpka}, we can see how our method compares to the cross-correlation method and the algorithm described in~\cite{Z_Hydro_Cirpka2007} for various $\textbf{y}$ input SNRs and 1000 and 5000 data points respectively (positive time interval of residence time to be estimated of 500 data points). Our method and the~\cite{Z_Hydro_Cirpka2007} algorithm show similarly good results in comparison with the cross-correlation. The method of~\cite{Z_Hydro_Cirpka2007} has a smaller standard deviation than our method, showing a weaker dependence of the noise/structure of the dataset.

While our proposed approach  provides different output results depending on the given $\lambda$, the best solution being picked automatically, the operator can choose an appropriate solution based on his own expertise, from an appropriate range around the optimal $\lambda$. Moreover, the solution is independent from the initialization due to the convexity of the $J$ functional. 

\begin{figure}[H]
    \centering
    \subfigure[]{\label{subfig_24} \includegraphics[trim=2cm 18cm 2cm 3cm, scale=0.55]{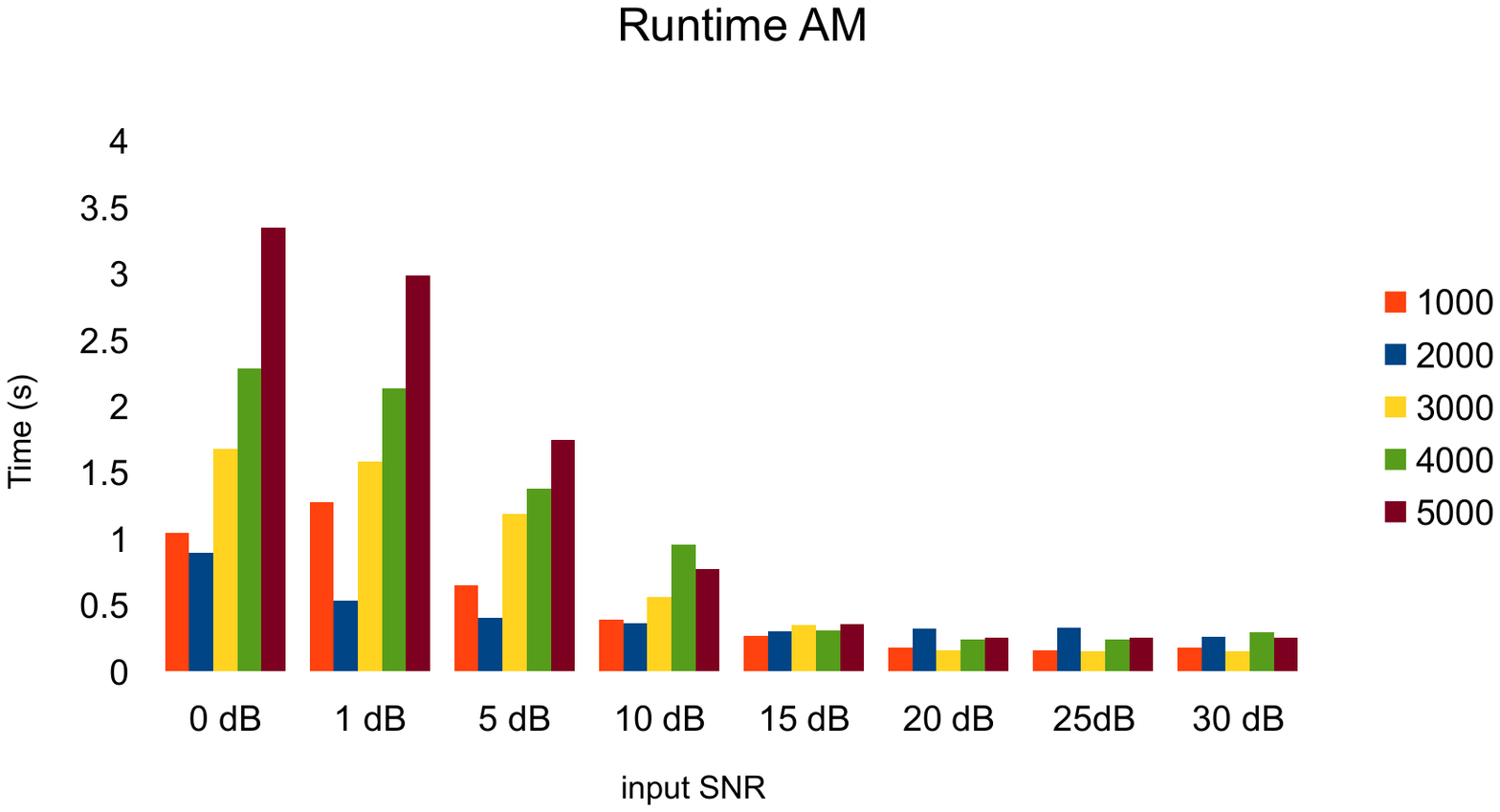}}
    \subfigure[]{\label{subfig_25} \includegraphics[trim=2cm 18cm 2cm 3cm, scale=0.55]{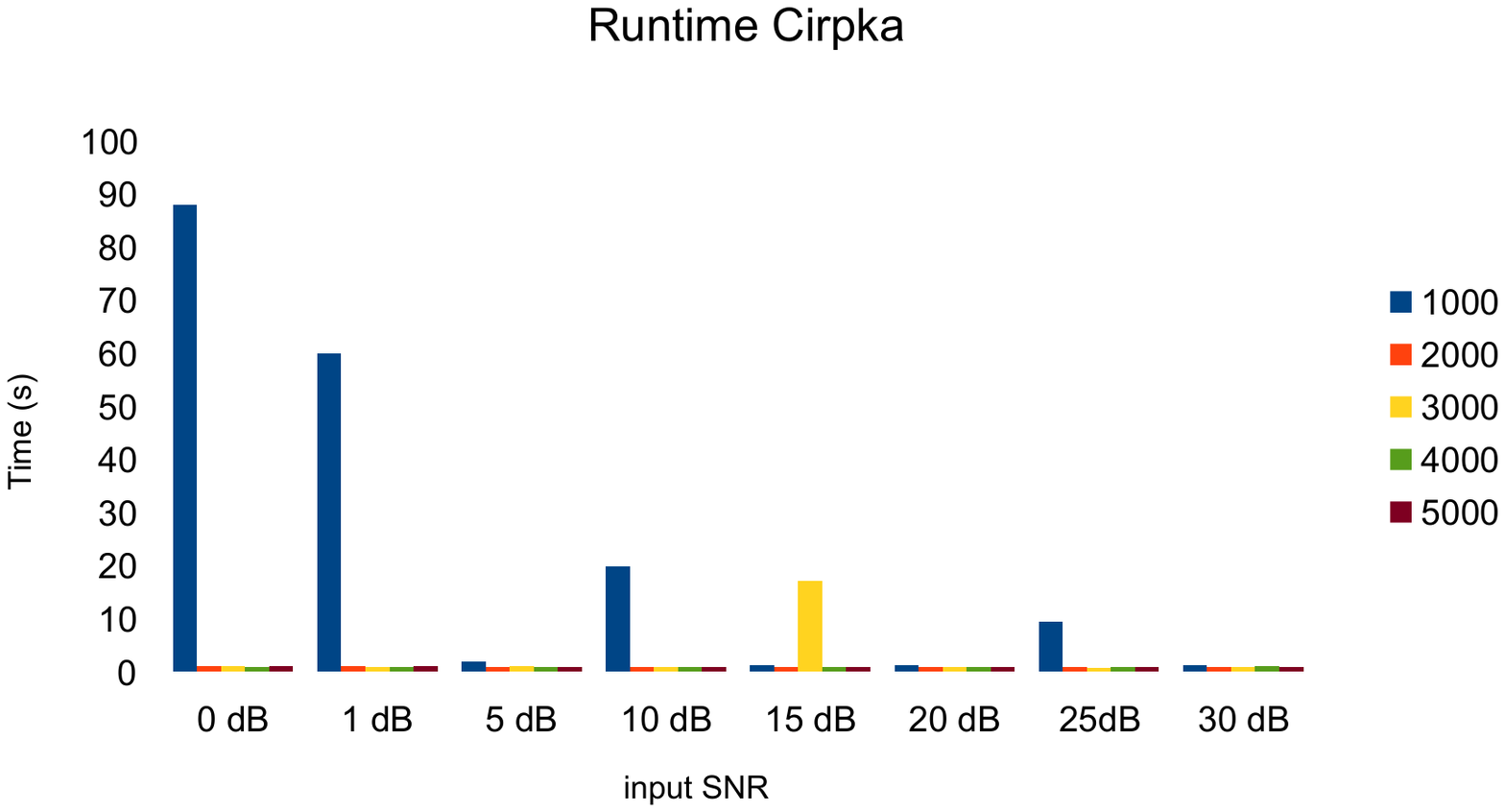}}
    \subfigure[]{\label{subfig_26} \includegraphics[trim=2cm 18cm 2cm 2.5cm, scale=0.55]{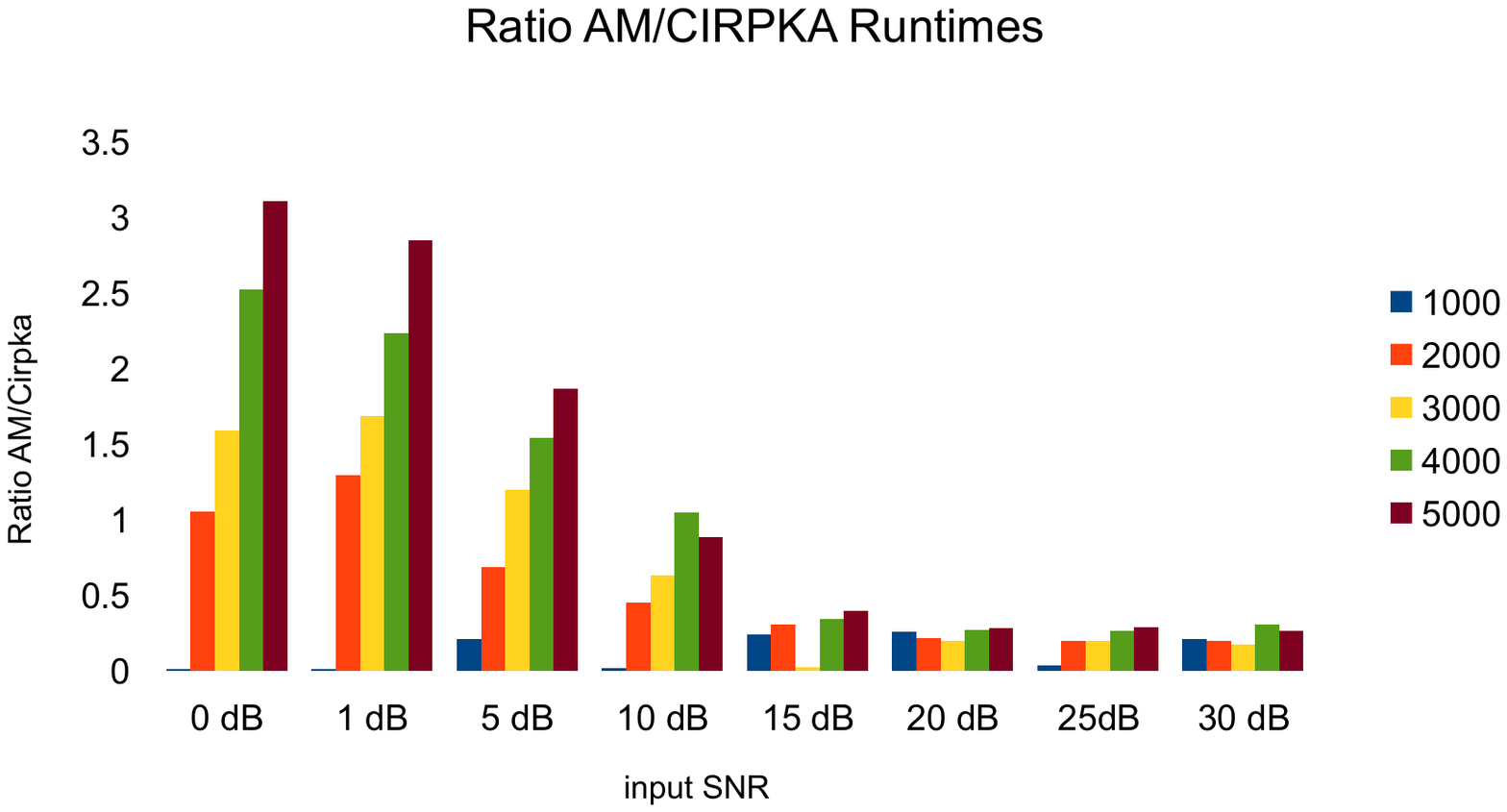}}
    \caption{Analysis of runtimes between the AM algorithm and the~\cite{Z_Hydro_Cirpka2007} algorithm for various lengths of the dataset and various noise levels.}
    \label{fig_runtimes_AM_cirpka}
\end{figure}

In Figure~\ref{fig_runtimes_AM_cirpka}, bar plots  illustrate the average runtime for 30 test cases, for different $\textbf{y}$ input SNRs, for the three algorithms.
The AM algorithm is consistently faster than the ~\cite{Z_Hydro_Cirpka2007} algorithm for $\textbf{y}$ input SNRs higher than 15 dB \ref{subfig_26}. It is also faster for the small data sets of 1000 points \ref{subfig_24},\ref{subfig_25}.
\\

\section{Real Data}
\label{sec:real}

The tests on real data are conducted on data sets made available from the "Base de Données des Observatoires en Hydrologie" $\copyright$ Irstea,~\cite{IRSTEA}. The data is gathered in the Ile de France region, in France. The measurements are from two neighboring sites, one at a higher altitude for rainfall measurements and the second at a lower altitude for aquifer measurements, taken at every 1 hour intervals, between January $1^{st}$, 2016 until January $1^{st}$, 2017.

\begin{figure}[H]
    \centering
    \subfigure[]{\label{subfig_27}\includegraphics[clip, trim=2cm 2cm 3cm 1cm, scale = 0.4]{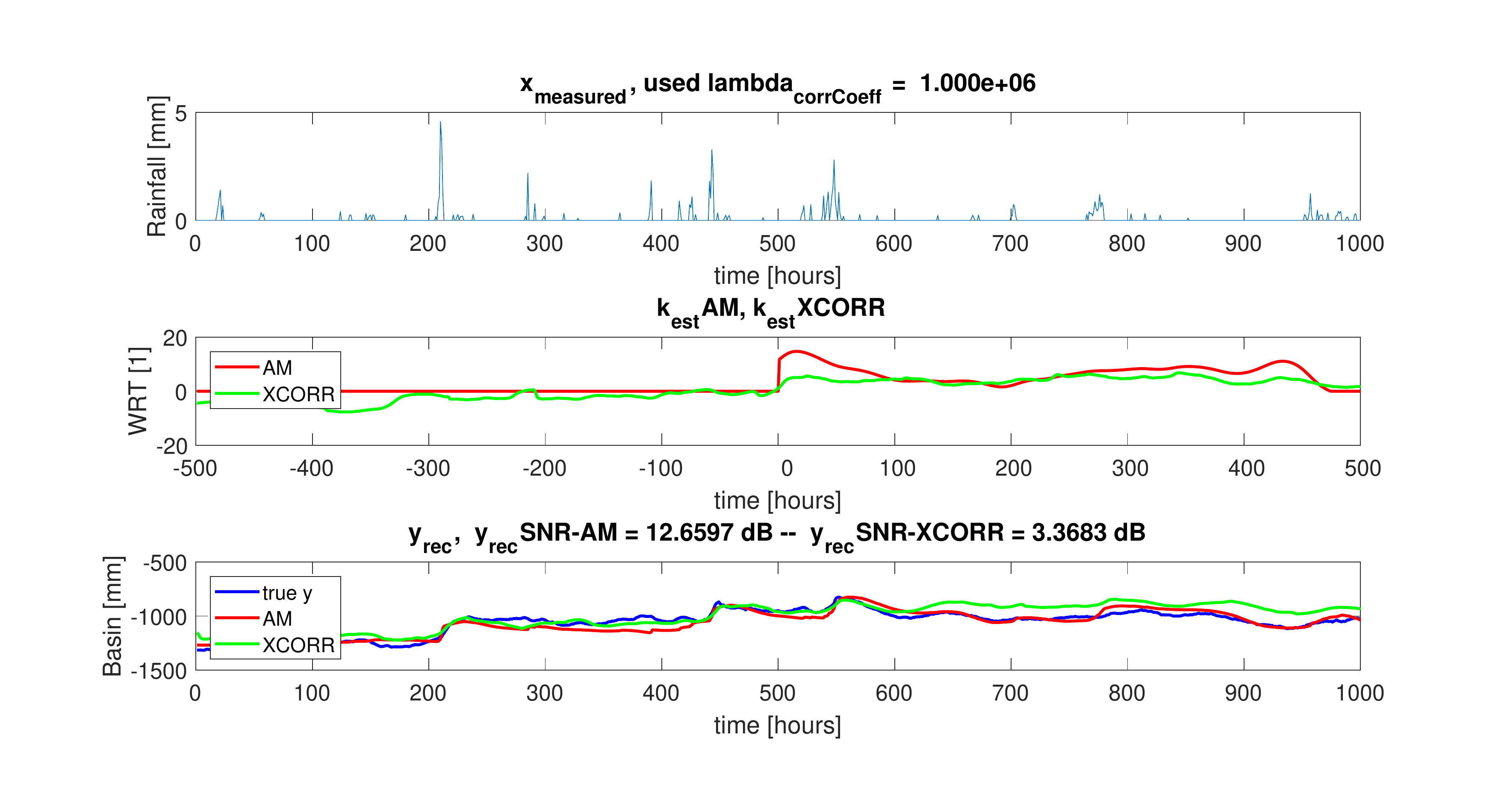}}
    \subfigure[]{\label{subfig_28}\includegraphics[clip, trim=2cm 2cm 2cm 1cm, scale = 0.4]{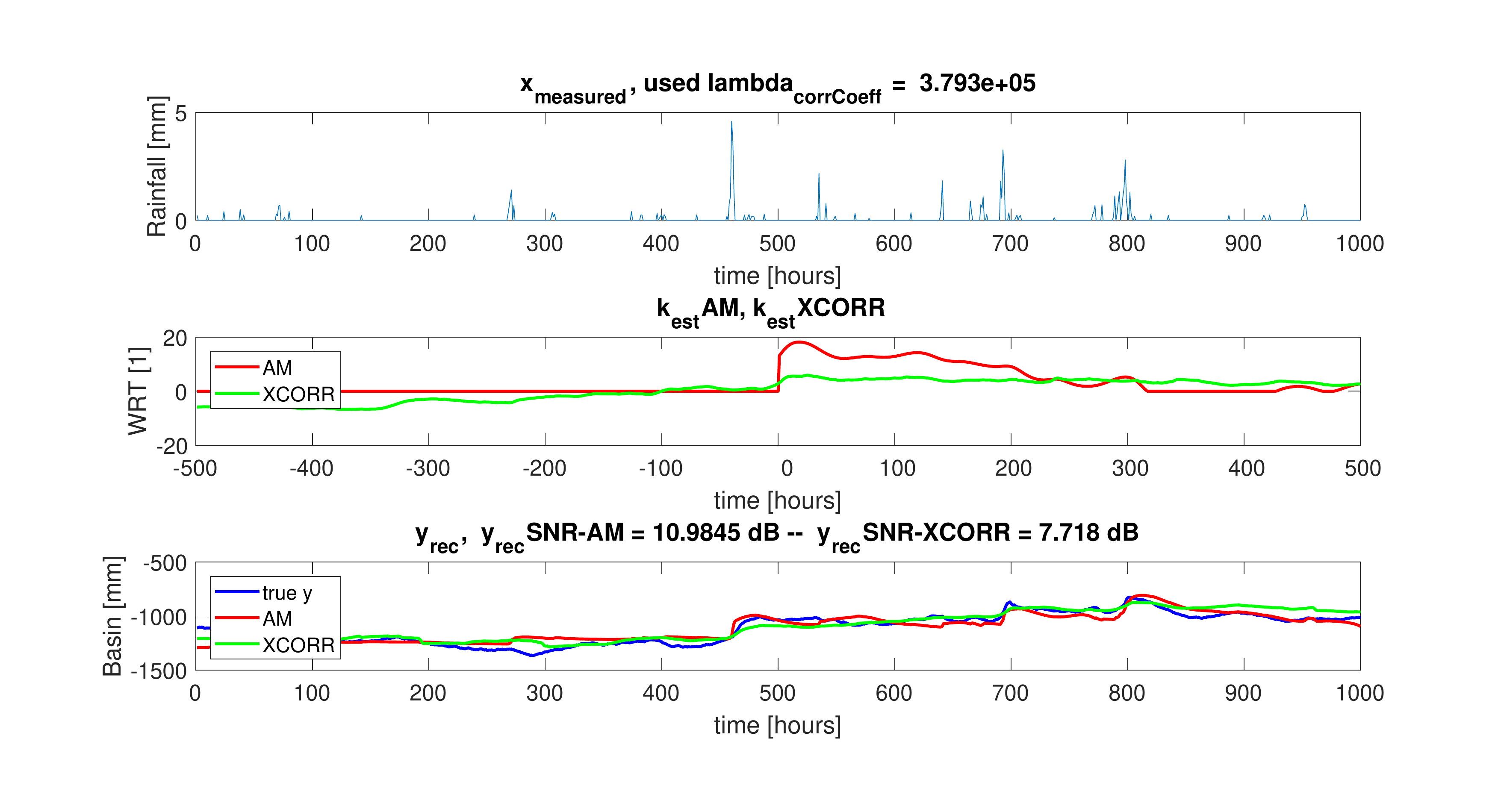}}
    \caption{Examples of results for real data using the  $\lambda_{corrCoeff}$ strategy. We estimate the residence time $\textbf{k}_{est}$ and the aquifer level $\textbf{c}_{est}$; we also plot the breakthrough curve $\textbf{y}_{rec}$ in blue. AM stands for the Alternating Minimization, XCORR for the standard cross-correlation, the true residence time $\textbf{k}$ is not known. The position of the maximum amplitude of $\textbf{k}_{est}$ is similar for the two methods but the shape of $\textbf{k}_{est}$ varies significantly. Only the AM method has the physical properties of positivity and causality.}
    \label{fig_real}
\end{figure}

\begin{figure}[H]
    \centering
    \subfigure[]{\label{subfig_29}\includegraphics[clip, trim=2cm 2cm 3cm 1cm, scale = 0.4]{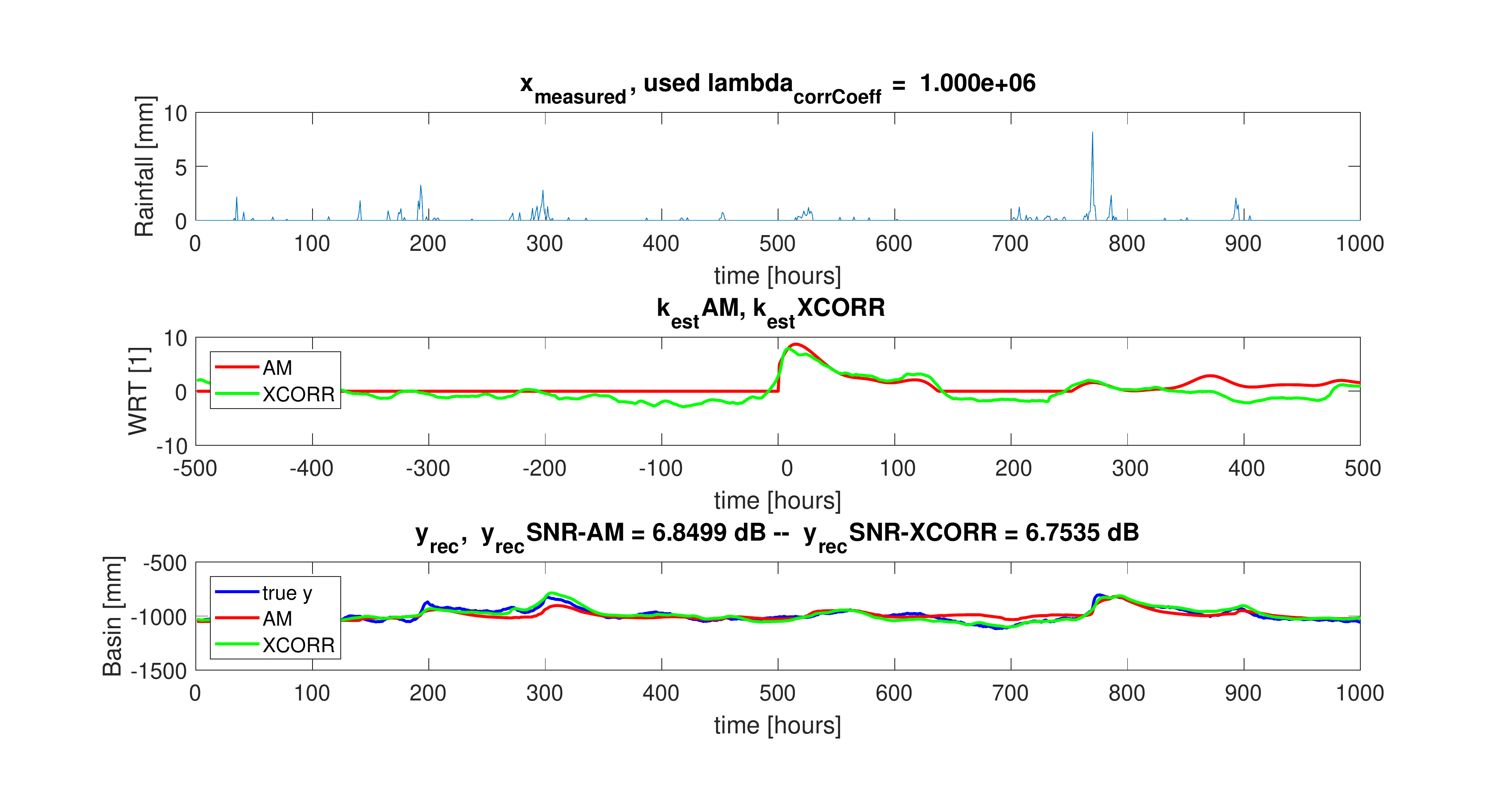}}
    \subfigure[]{\label{subfig_30}\includegraphics[clip, trim=2cm 2cm 3cm 1cm, scale = 0.4]{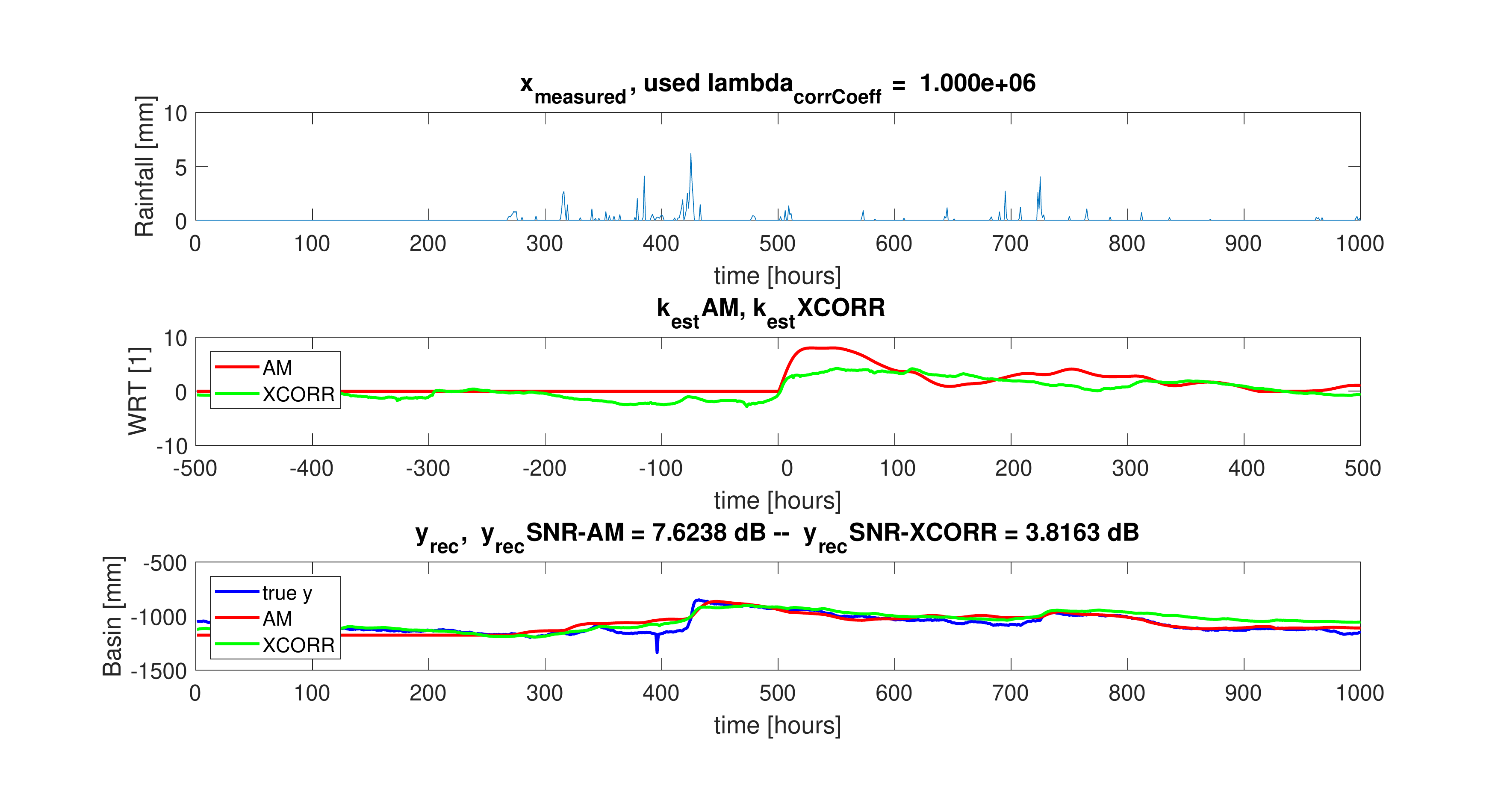}}
    \caption{Same as in Figure~\ref{fig_real}.}
    \label{fig_real_2}
\end{figure}

For the real data, the estimates are based on the  $\lambda_{corrCoeff}$ strategy with $\lambda$s chosen around the optimal values found with the synthetic data set, between $10^{8}$ to $10^{2}$.
In Figure~\ref{fig_real} and in Figure~\ref{fig_real_2}, estimates of the residence time for real life measurements of $\textbf{x}$ and $\textbf{y}$ are shown. 

In all cases, the estimated curves honor the given positivity and causality constraints. For the cross-correlation, even if the $\textbf {y}_{rec}$ is close to the original $\textbf{y}$, the curve for the residence time estimated by this method has the disadvantage to not respect the positivity and causality constraints across all of the presented cases.

The aquifer level measurements have negative values due to the conventions of the used measuring instruments. The AM algorithm is also capable of estimating the aquifer average level $\textbf{c}$, and depending on this constant and the amplitude of the rain fall input, the estimated residence time curve $\textbf{k}_{est}$ will also have a certain amplitude (the curve is not normalized to resemble that of a pdf).

The AM algorithm succeeds in reconstructing the $\textbf{y}_{rec}$ with an SNR around 10 dB in the studied cases, using the $\lambda_{corrCoeff}$ and provides a better reconstruction SNR than the cross-correlation (XCORR) method. 

We find small but significant changes in the residence time curve for different data sets of the same channel, as also identified in other datasets~\cite{Z_Hydro_Delbart_2014}. This may be due to the seasonal variability of the inputs (rainfall) and its effects on the hydrological process. This aspect would be of interest to study into more detail for specific sites to better understand it.

Another observation to be made is the fact that if non-linearities of the system are present (in transit or at the aquifer level), our approach may also lead to over simplification. Nonetheless the question arises if a hydrological channel could be considered as a linear and stationary system by parts (smaller time series) and therefore allow the use of our method for estimating partial residence time curves which can then be put together in a more complex mapping of the channel.

One can also note in the plots that the $\textbf{y}_{rec}$ is slightly better for cases when a heavy rainfall event appears at the beginning of the time series $\textbf{x}$ instead of towards the end, suggesting the fact that the residence time estimation would also be better.

Finally, the examples show the appearance of multiple lobes that are considered a sign of reservoirs of the hydrological channel keeping part of the water for some time before releasing it in a later discharge. This demonstrates the usefulness of a non-parametric deconvolution method in comparison with parametric deconvolution methods where such lobes are either ignored or fixed in number.

\section{Conclusion}
\label{sec:conclusion}

We propose a new approach to estimate a smooth residence time taking into account positivity and causality constraints and having a fast runtime. We highlight why these constraints must be used all along the algorithmic process to reach the expected solution in the case of non-parametric 1D deconvolution for the AM algorithm presented here.

The estimation of the residence time $\textbf{k}_{est}$ was done using a fast Alternating Minimization algorithm with two steps: (1) 1D deconvolution and (2) estimation of the aquifer initial level. All tests have been done on a personal laptop, with CPU Intel(R) Core(TM) i7-6600U CPU @ 2.6GHz 2.81 GHz, 16.0 GB RAM, 64-bit OS, x-64-based processor, using Matlab\textsuperscript{\tiny\textregistered}. We validated the approach on synthetic tests and proposed several strategies to automatically estimate a hyper-parameter, $\lambda$, that controls the smoothness of the residence time curve. We have found that between these strategies, the correlation coefficient strategy seems to be very efficient to estimate the best value for $\lambda$.

We validated our AM method on synthetic data and found that the results are better than the standard cross-correlation method and similar to those of the \cite{Z_Hydro_Cirpka2007} method. We also demonstrated the capabilities of our AM method on real data. Additionally, our method respects the physical constraints (positivity, causality, non-circularity) which are important for interpretation purposes. The estimation made by our method will provide better information for hydro-geologists on amplitude and full shape of the residence time, the mean level of the aquifer and will also improve the estimation of the mean residence time (\ref{app:B} shows how to compute it).

As possible improvements we propose refining this methodology for the potential non-linear aspects of the water transit time through the medium.

The Matlab implementation of the algorithm is available under CECILL license at the following public Git repository: \url{https://git.l2s.centralesupelec.fr/meresescual/SmoothSignalEstimatorDeconvolution.git}.

\section*{Acknowledgments}

We thank the \textit{Base de Données des Observatoires en Hydrologie} for providing the data acquired in the field.
$\copyright$  Irstea, BDOH$\_$ORACLE$\copyright$, July 24, 2017. 

We thank Prof. Olaf A. Cirpka at the department for Hydrogeology, Universt{\"a}t T{\"u}bingen, Germany, for kindly providing the algorithm referenced in~\cite{Z_Hydro_Cirpka2007}.

We thank Amine Hadjyoucef and Christian Kuschel for carefully proof-reading this text and offering us their comments about unclarities and English phrasing.

We also want to thank the editor and the reviewers for their constructive questions and observations throughout the reviewing process.

This work is supported by the Center for Data Science, funded by the IDEX Paris-Saclay, ANR-11-IDEX-0003-02. We acknowledge support from the "Institut National des Sciences de l'Univers" (INSU), the "Centre National de la Recherche Scientifique" (CNRS) and "Centre National d'Etude Spatiale" (CNES) and through the "Programme National de Planétologie" and MEX/PFS Program.


\appendix

\section{Mean Residence Time}
\label{app:B}

In order to estimate the mean residence time $\tau$, one has to simply renormalize the estimated transfer function $\textbf{k}_{est}$ and take the mean:
\begin{equation}\label{21}
\begin{aligned}
& \tau = \displaystyle\sum_{t=0}^{t = \dfrac{K}{2}} \left( \dfrac{\textbf{k}_{est}(t) \cdot t }{\displaystyle\sum_{t=0}^{t = \dfrac{K}{2}} \textbf{k}_{est}(t)} \right)
\end{aligned}
\end{equation}

\section*{References}

\bibliography{mybibfile}

\begin{thebibliography}{31}
\expandafter\ifx\csname natexlab\endcsname\relax\def\natexlab#1{#1}\fi
\providecommand{\url}[1]{\texttt{#1}}
\providecommand{\href}[2]{#2}
\providecommand{\path}[1]{#1}
\providecommand{\DOIprefix}{doi:}
\providecommand{\ArXivprefix}{arXiv:}
\providecommand{\URLprefix}{URL: }
\providecommand{\Pubmedprefix}{pmid:}
\providecommand{\doi}[1]{\href{http://dx.doi.org/#1}{\path{#1}}}
\providecommand{\Pubmed}[1]{\href{pmid:#1}{\path{#1}}}
\providecommand{\bibinfo}[2]{#2}
\ifx\xfnm\relax \def\xfnm[#1]{\unskip,\space#1}\fi
\bibitem[{Bertsekas(1982)}]{Z_Hydro_bertsekas1982}
\bibinfo{author}{Bertsekas, D.~P.} (\bibinfo{year}{1982}).
\newblock \bibinfo{title}{Projected newton methods for optimization problems
  with simple constraints}.
\newblock {\it \bibinfo{journal}{SIAM Journal on control and Optimization}\/},
  {\it \bibinfo{volume}{20}\/}, \bibinfo{pages}{221--246}.
  \DOIprefix\doi{https://doi.org/10.1137/0320018}.
\bibitem[{Botter et~al.(2011)Botter, Bertuzzo \& Rinaldo}]{Z_Hydro_Botter2011}
\bibinfo{author}{Botter, G.}, \bibinfo{author}{Bertuzzo, E.}, \&
  \bibinfo{author}{Rinaldo, A.} (\bibinfo{year}{2011}).
\newblock \bibinfo{title}{Catchment residence and travel time distributions:
  The master equation}.
\newblock {\it \bibinfo{journal}{Geophysical Research Letters}\/},  {\it
  \bibinfo{volume}{38}\/}, \bibinfo{pages}{n/a--n/a}. \URLprefix
  \url{http://dx.doi.org/10.1029/2011GL047666}.
  \DOIprefix\doi{10.1029/2011GL047666}.
\newblock \bibinfo{note}{L11403}.
\bibitem[{Cirpka et~al.(2007)Cirpka, Fienen, Hofer, Hoehn, Tessarini, Kipfer \&
  Kitanidis}]{Z_Hydro_Cirpka2007}
\bibinfo{author}{Cirpka, O.~A.}, \bibinfo{author}{Fienen, M.~N.},
  \bibinfo{author}{Hofer, M.}, \bibinfo{author}{Hoehn, E.},
  \bibinfo{author}{Tessarini, A.}, \bibinfo{author}{Kipfer, R.}, \&
  \bibinfo{author}{Kitanidis, P.~K.} (\bibinfo{year}{2007}).
\newblock \bibinfo{title}{Analyzing bank filtration by deconvoluting time
  series of electric conductivity}.
\newblock {\it \bibinfo{journal}{Ground Water}\/},  {\it
  \bibinfo{volume}{45}\/}, \bibinfo{pages}{318--328}. \URLprefix
  \url{http://dx.doi.org/10.1111/j.1745-6584.2006.00293.x}.
  \DOIprefix\doi{10.1111/j.1745-6584.2006.00293.x}.
\bibitem[{Delbart et~al.(2014)Delbart, Valdes, Barbecot, Tognelli, Richon \&
  Couchoux}]{Z_Hydro_Delbart_2014}
\bibinfo{author}{Delbart, C.}, \bibinfo{author}{Valdes, D.},
  \bibinfo{author}{Barbecot, F.}, \bibinfo{author}{Tognelli, A.},
  \bibinfo{author}{Richon, P.}, \& \bibinfo{author}{Couchoux, L.}
  (\bibinfo{year}{2014}).
\newblock \bibinfo{title}{Temporal variability of karst aquifer response time
  established by the sliding-windows cross-correlation method}.
\newblock {\it \bibinfo{journal}{Journal of Hydrology}\/},  {\it
  \bibinfo{volume}{511}\/}, \bibinfo{pages}{580--588}.
  \DOIprefix\doi{10.1016/j.jhydrol.2014.02.008}.
\bibitem[{Dietrich \& Chapman(1993)}]{Dietrich1993}
\bibinfo{author}{Dietrich, C.}, \& \bibinfo{author}{Chapman, T.}
  (\bibinfo{year}{1993}).
\newblock \bibinfo{title}{Unit graph estimation and stabilization using
  quadratic programming and difference norms}.
\newblock {\it \bibinfo{journal}{Water resources research}\/},  {\it
  \bibinfo{volume}{29}\/}, \bibinfo{pages}{2629--2635}.
\bibitem[{Dzikowski \& Delay(1992)}]{Z_Hydro_DZIKOWSKI1992697}
\bibinfo{author}{Dzikowski, M.}, \& \bibinfo{author}{Delay, F.}
  (\bibinfo{year}{1992}).
\newblock \bibinfo{title}{Simulation algorithm of time-dependent tracer test
  systems in hydrogeology}.
\newblock {\it \bibinfo{journal}{Computers \& Geosciences}\/},  {\it
  \bibinfo{volume}{18}\/}, \bibinfo{pages}{697 -- 705}. \URLprefix
  \url{http://www.sciencedirect.com/science/article/pii/009830049290004B}.
  \DOIprefix\doi{http://dx.doi.org/10.1016/0098-3004(92)90004-B}.
\bibitem[{Etcheverry \& Perrochet(2000)}]{Z_Hydro_Etcheverry2000}
\bibinfo{author}{Etcheverry, D.}, \& \bibinfo{author}{Perrochet, P.}
  (\bibinfo{year}{2000}).
\newblock \bibinfo{title}{Direct simulation of groundwater transit-time
  distributions using the reservoir theory}.
\newblock {\it \bibinfo{journal}{Hydrogeology Journal}\/},  {\it
  \bibinfo{volume}{8}\/}, \bibinfo{pages}{200--208}. \URLprefix
  \url{http://dx.doi.org/10.1007/s100400050006}.
  \DOIprefix\doi{10.1007/s100400050006}.
\bibitem[{Fienen et~al.(2008)Fienen, Clemo \& Kitanidis}]{Fienen2008}
\bibinfo{author}{Fienen, M.~N.}, \bibinfo{author}{Clemo, T.}, \&
  \bibinfo{author}{Kitanidis, P.~K.} (\bibinfo{year}{2008}).
\newblock \bibinfo{title}{An interactive bayesian geostatistical inverse
  protocol for hydraulic tomography}.
\newblock {\it \bibinfo{journal}{Water Resources Research}\/},  {\it
  \bibinfo{volume}{44}\/}.
\bibitem[{Fienen et~al.(2006)Fienen, Luo \& Kitanidis}]{Fienen2006}
\bibinfo{author}{Fienen, M.~N.}, \bibinfo{author}{Luo, J.}, \&
  \bibinfo{author}{Kitanidis, P.~K.} (\bibinfo{year}{2006}).
\newblock \bibinfo{title}{A bayesian geostatistical transfer function approach
  to tracer test analysis}.
\newblock {\it \bibinfo{journal}{Water Resources Research}\/},  {\it
  \bibinfo{volume}{42}\/}.
\bibitem[{Gooseff et~al.(2011)Gooseff, Benson, Briggs, Weaver, Wollheim,
  Peterson \& Hopkinson}]{Z_Hydro_Gooseff2011}
\bibinfo{author}{Gooseff, M.~N.}, \bibinfo{author}{Benson, D.~A.},
  \bibinfo{author}{Briggs, M.~A.}, \bibinfo{author}{Weaver, M.},
  \bibinfo{author}{Wollheim, W.}, \bibinfo{author}{Peterson, B.}, \&
  \bibinfo{author}{Hopkinson, C.~S.} (\bibinfo{year}{2011}).
\newblock \bibinfo{title}{Residence time distributions in surface transient
  storage zones in streams: Estimation via signal deconvolution}.
\newblock {\it \bibinfo{journal}{Water Resources Research}\/},  {\it
  \bibinfo{volume}{47}\/}, \bibinfo{pages}{n/a--n/a}. \URLprefix
  \url{http://dx.doi.org/10.1029/2010WR009959}.
  \DOIprefix\doi{10.1029/2010WR009959}.
\newblock \bibinfo{note}{W05509}.
\bibitem[{Hoehn \& Cirpka(2006)}]{HoehnCirpka2006}
\bibinfo{author}{Hoehn, E.}, \& \bibinfo{author}{Cirpka, O.~A.}
  (\bibinfo{year}{2006}).
\newblock \bibinfo{title}{Assessing residence times of hyporheic ground water
  in two alluvial flood plains of the southern alps using water temperature and
  tracers}.
\newblock {\it \bibinfo{journal}{Hydrology and Earth System Sciences}\/},  {\it
  \bibinfo{volume}{10}\/}, \bibinfo{pages}{553--563}. \URLprefix
  \url{https://www.hydrol-earth-syst-sci.net/10/553/2006/}.
  \DOIprefix\doi{10.5194/hess-10-553-2006}.
\bibitem[{Irstea(2017)}]{IRSTEA}
\bibinfo{author}{Irstea} (\bibinfo{year}{2017}).
\newblock \bibinfo{title}{"base de données des observatoires en hydrologie"
  $\copyright$ irstea}.
\newblock \URLprefix \url{https://bdoh.irstea.fr/ORACLE/}.
\bibitem[{Long \& Derickson(1999)}]{Long1999}
\bibinfo{author}{Long, A.}, \& \bibinfo{author}{Derickson, R.}
  (\bibinfo{year}{1999}).
\newblock \bibinfo{title}{Linear systems analysis in a karst aquifer}.
\newblock {\it \bibinfo{journal}{Journal of Hydrology}\/},  {\it
  \bibinfo{volume}{219}\/}, \bibinfo{pages}{206--217}.
\bibitem[{Luo et~al.(2006)Luo, Cirpka, Fienen, Wu, Mehlhorn, Carley, Jardine,
  Criddle \& Kitanidis}]{Luo2006}
\bibinfo{author}{Luo, J.}, \bibinfo{author}{Cirpka, O.~A.},
  \bibinfo{author}{Fienen, M.~N.}, \bibinfo{author}{Wu, W.-m.},
  \bibinfo{author}{Mehlhorn, T.~L.}, \bibinfo{author}{Carley, J.},
  \bibinfo{author}{Jardine, P.~M.}, \bibinfo{author}{Criddle, C.~S.}, \&
  \bibinfo{author}{Kitanidis, P.~K.} (\bibinfo{year}{2006}).
\newblock \bibinfo{title}{A parametric transfer function methodology for
  analyzing reactive transport in nonuniform flow}.
\newblock {\it \bibinfo{journal}{Journal of contaminant hydrology}\/},  {\it
  \bibinfo{volume}{83}\/}, \bibinfo{pages}{27--41}.
\bibitem[{Massei et~al.(2006)Massei, Dupont, Mahler, Laignel, Fournier, Valdes
  \& Ogier}]{Z_Hydro_Massei2006}
\bibinfo{author}{Massei, N.}, \bibinfo{author}{Dupont, J.},
  \bibinfo{author}{Mahler, B.}, \bibinfo{author}{Laignel, B.},
  \bibinfo{author}{Fournier, M.}, \bibinfo{author}{Valdes, D.}, \&
  \bibinfo{author}{Ogier, S.} (\bibinfo{year}{2006}).
\newblock \bibinfo{title}{Investigating transport properties and turbidity
  dynamics of a karst aquifer using correlation, spectral, and wavelet
  analyses}.
\newblock {\it \bibinfo{journal}{Journal of Hydrology}\/},  {\it
  \bibinfo{volume}{329}\/}, \bibinfo{pages}{244 -- 257}. \URLprefix
  \url{http://www.sciencedirect.com/science/article/pii/S0022169406000965}.
  \DOIprefix\doi{http://doi.org/10.1016/j.jhydrol.2006.02.021}.
\bibitem[{McCormick(1969)}]{McCormick1969}
\bibinfo{author}{McCormick, G.~P.} (\bibinfo{year}{1969}).
\newblock \bibinfo{title}{Anti-zig-zagging by bending}.
\newblock {\it \bibinfo{journal}{Management Science}\/},  (pp.
  \bibinfo{pages}{315--320}).
\bibitem[{McGuire \& McDonnell(2006)}]{Z_Hydro_McGuireMcDonnell2006}
\bibinfo{author}{McGuire, K.~J.}, \& \bibinfo{author}{McDonnell, J.~J.}
  (\bibinfo{year}{2006}).
\newblock \bibinfo{title}{A review and evaluation of catchment transit time
  modeling}.
\newblock {\it \bibinfo{journal}{Journal of Hydrology}\/},  {\it
  \bibinfo{volume}{330}\/}, \bibinfo{pages}{543--563}. \URLprefix
  \url{http://dx.doi.org/10.1016/j.jhydrol.2006.04.020}.
  \DOIprefix\doi{10.1016/j.jhydrol.2006.04.020}.
\bibitem[{Michalak \& Kitanidis(2003)}]{Michalak2003}
\bibinfo{author}{Michalak, A.~M.}, \& \bibinfo{author}{Kitanidis, P.~K.}
  (\bibinfo{year}{2003}).
\newblock \bibinfo{title}{A method for enforcing parameter nonnegativity in
  bayesian inverse problems with an application to contaminant source
  identification}.
\newblock {\it \bibinfo{journal}{Water Resources Research}\/},  {\it
  \bibinfo{volume}{39}\/}.
\bibitem[{Neuman \& De~Marsily(1976)}]{Z_Hydro_NeumanDeMarsily1976}
\bibinfo{author}{Neuman, S.~P.}, \& \bibinfo{author}{De~Marsily, G.}
  (\bibinfo{year}{1976}).
\newblock \bibinfo{title}{Identification of linear systems response by
  parametric programing}.
\newblock {\it \bibinfo{journal}{Water Resources Research}\/},  {\it
  \bibinfo{volume}{12}\/}, \bibinfo{pages}{253--262}. \URLprefix
  \url{http://dx.doi.org/10.1029/WR012i002p00253}.
  \DOIprefix\doi{10.1029/WR012i002p00253}.
\bibitem[{Neuman et~al.(1982)Neuman, Resnick, Reebles \&
  Dunbar}]{Z_Hydro_Neuman1982}
\bibinfo{author}{Neuman, S.~P.}, \bibinfo{author}{Resnick, S.~D.},
  \bibinfo{author}{Reebles, R.~W.}, \& \bibinfo{author}{Dunbar, D.~B.}
  (\bibinfo{year}{1982}).
\newblock \bibinfo{title}{Developing a new deconvolution technique to model
  rainfall-runoff in arid environments}.
\newblock {\it \bibinfo{journal}{Water Resources Research Center, University of
  Arizona}\/}, . \URLprefix \url{http://hdl.handle.net/10150/305311}.
\bibitem[{O’Sullivan(1998)}]{Z_Hydro_O_Sullivan_1998}
\bibinfo{author}{O’Sullivan, J.~A.} (\bibinfo{year}{1998}).
\newblock \bibinfo{title}{Alternating minimization algorithms: From
  blahut-arimoto to expectation-maximization}.
\newblock {\it \bibinfo{journal}{Springer Science+Business Media New York}\/},
  (pp. \bibinfo{pages}{173--192}).
  \DOIprefix\doi{10.1007/978-1-4615-5121-8_13}.
\bibitem[{Payn et~al.(2008)Payn, Gooseff, Benson, Cirpka, Zarnetske, Bowden,
  McNamara \& Bradford}]{Z_Hydro_PaynGooseff2008}
\bibinfo{author}{Payn, R.~A.}, \bibinfo{author}{Gooseff, M.~N.},
  \bibinfo{author}{Benson, D.~A.}, \bibinfo{author}{Cirpka, O.~A.},
  \bibinfo{author}{Zarnetske, J.~P.}, \bibinfo{author}{Bowden, W.~B.},
  \bibinfo{author}{McNamara, J.~P.}, \& \bibinfo{author}{Bradford, J.~H.}
  (\bibinfo{year}{2008}).
\newblock \bibinfo{title}{Comparison of instantaneous and constant-rate stream
  tracer experiments through non-parametric analysis of residence time
  distributions}.
\newblock {\it \bibinfo{journal}{Water Resources Research}\/},  {\it
  \bibinfo{volume}{44}\/}, \bibinfo{pages}{n/a--n/a}. \URLprefix
  \url{http://dx.doi.org/10.1029/2007WR006274}.
  \DOIprefix\doi{10.1029/2007WR006274}.
\newblock \bibinfo{note}{W06404}.
\bibitem[{Pereverzev \& Schock(2009)}]{Z_Hydro_Pereverzev2009}
\bibinfo{author}{Pereverzev, S.}, \& \bibinfo{author}{Schock, E.}
  (\bibinfo{year}{2009}).
\newblock \bibinfo{title}{Morozov's discrepancy principle for tikhonov
  regularization of severely ill-posed problems in finite-dimensional
  subspaces.}
\newblock {\it \bibinfo{journal}{Numerical Functional Analysis and
  Optimization}\/}, .
  \DOIprefix\doi{http://dx.doi.org/10.1080/01630560008816993}.
\bibitem[{Provencher(1982)}]{Provencher1982a}
\bibinfo{author}{Provencher, S.~W.} (\bibinfo{year}{1982}).
\newblock \bibinfo{title}{Contin: a general purpose constrained regularization
  program for inverting noisy linear algebraic and integral equations}.
\newblock {\it \bibinfo{journal}{Computer Physics Communications}\/},  {\it
  \bibinfo{volume}{27}\/}, \bibinfo{pages}{229--242}.
\bibitem[{Robinson et~al.(2010)Robinson, Dash \&
  Srinivasan}]{Z_Hydro_Robinson2010}
\bibinfo{author}{Robinson, B.~A.}, \bibinfo{author}{Dash, Z.~V.}, \&
  \bibinfo{author}{Srinivasan, G.} (\bibinfo{year}{2010}).
\newblock \bibinfo{title}{A particle tracking transport method for the
  simulation of resident and flux-averaged concentration of solute plumes in
  groundwater models}.
\newblock {\it \bibinfo{journal}{Computational Geosciences}\/},  {\it
  \bibinfo{volume}{14}\/}, \bibinfo{pages}{779--792}. \URLprefix
  \url{http://dx.doi.org/10.1007/s10596-010-9190-6}.
  \DOIprefix\doi{10.1007/s10596-010-9190-6}.
\bibitem[{Sheets et~al.(2002)Sheets, Darner \& Whitteberry}]{Sheets2002}
\bibinfo{author}{Sheets, R.}, \bibinfo{author}{Darner, R.}, \&
  \bibinfo{author}{Whitteberry, B.} (\bibinfo{year}{2002}).
\newblock \bibinfo{title}{Lag times of bank filtration at a well field,
  cincinnati, ohio, usa}.
\newblock {\it \bibinfo{journal}{Journal of Hydrology}\/},  {\it
  \bibinfo{volume}{266}\/}, \bibinfo{pages}{162 -- 174}. \URLprefix
  \url{http://www.sciencedirect.com/science/article/pii/S0022169402001646}.
  \DOIprefix\doi{https://doi.org/10.1016/S0022-1694(02)00164-6}.
\newblock \bibinfo{note}{Attenuation of Groundwater Pollution by Bank
  Filtration}.
\bibitem[{Skaggs et~al.(1998)Skaggs, Kabala \& Jury}]{Skaggs1998}
\bibinfo{author}{Skaggs, T.~H.}, \bibinfo{author}{Kabala, Z.}, \&
  \bibinfo{author}{Jury, W.~A.} (\bibinfo{year}{1998}).
\newblock \bibinfo{title}{Deconvolution of a nonparametric transfer function
  for solute transport in soils}.
\newblock {\it \bibinfo{journal}{Journal of Hydrology}\/},  {\it
  \bibinfo{volume}{207}\/}, \bibinfo{pages}{170--178}.
\bibitem[{Tessier et~al.(1996)Tessier, Lovejoy, Hubert, Schertzer \&
  Pecknold}]{Z_Hydro_Tessier1996b}
\bibinfo{author}{Tessier, Y.}, \bibinfo{author}{Lovejoy, S.},
  \bibinfo{author}{Hubert, P.}, \bibinfo{author}{Schertzer, D.}, \&
  \bibinfo{author}{Pecknold, S.} (\bibinfo{year}{1996}).
\newblock \bibinfo{title}{Multifractal analysis and modeling of rainfall and
  river flows and scaling, causal transfer functions}.
\newblock {\it \bibinfo{journal}{Journal of Geophysical Research:
  Atmospheres}\/},  {\it \bibinfo{volume}{101}\/},
  \bibinfo{pages}{26427--26440}. \URLprefix
  \url{http://dx.doi.org/10.1029/96JD01799}. \DOIprefix\doi{10.1029/96JD01799}.
\bibitem[{Vogt et~al.(2010)Vogt, Hoehn, Schneider, Freund, Schirmer \&
  Cirpka}]{Vogt2010}
\bibinfo{author}{Vogt, T.}, \bibinfo{author}{Hoehn, E.},
  \bibinfo{author}{Schneider, P.}, \bibinfo{author}{Freund, A.},
  \bibinfo{author}{Schirmer, M.}, \& \bibinfo{author}{Cirpka, O.~A.}
  (\bibinfo{year}{2010}).
\newblock \bibinfo{title}{Fluctuations of electrical conductivity as a natural
  tracer for bank filtration in a losing stream}.
\newblock {\it \bibinfo{journal}{Advances in Water Resources}\/},  {\it
  \bibinfo{volume}{33}\/}, \bibinfo{pages}{1296--1308}.
\bibitem[{Werner \& Kadlec(2000)}]{Z_Hydro_WernerKadlec2000}
\bibinfo{author}{Werner, T.~M.}, \& \bibinfo{author}{Kadlec, R.~H.}
  (\bibinfo{year}{2000}).
\newblock \bibinfo{title}{Wetland residence time distribution modeling}.
\newblock {\it \bibinfo{journal}{Ecological Engineering}\/},  {\it
  \bibinfo{volume}{15}\/}, \bibinfo{pages}{77--90}. \URLprefix
  \url{http://dx.doi.org/10.1016/S0925-8574(99)00036-1}.
  \DOIprefix\doi{10.1016/s0925-8574(99)00036-1}.
\bibitem[{Zuo \& Hu(2012)}]{Z_Hydro_Zuo2012170}
\bibinfo{author}{Zuo, B.}, \& \bibinfo{author}{Hu, X.} (\bibinfo{year}{2012}).
\newblock \bibinfo{title}{Geophysical model enhancement technique based on
  blind deconvolution}.
\newblock {\it \bibinfo{journal}{Computers \& Geosciences}\/},  {\it
  \bibinfo{volume}{49}\/}, \bibinfo{pages}{170 -- 181}. \URLprefix
  \url{http://www.sciencedirect.com/science/article/pii/S0098300412002129}.
  \DOIprefix\doi{http://doi.org/10.1016/j.cageo.2012.06.017}.

\end{thebibliography}

\end{document}